\documentclass[]{jfm}

\usepackage{graphicx,caption}

\usepackage{amsmath}
\usepackage{amssymb}
\usepackage{newtxtext}
\usepackage{newtxmath,bm}
\usepackage{amsfonts}
\usepackage{color}
\usepackage{comment}

\newcommand{\Wi}{\mathrm{Wi}}
\renewcommand{\Re}{\mathrm{Re}}

\usepackage{natbib}
\newcommand{\RomanNumeralCaps}[1]
\linenumbers

\renewcommand{\c}{\mathsfbi{c}}

\shortauthor{Damiano Capocci, Moritz Linkmann, and Alexander Morozov}

\title{The unimportance of staying positive} 
\title{No need to stay positive: a practical approach to direct numerical simulations of elastic turbulence}

\author{Damiano Capocci\aff{1},
  Moritz Linkmann\aff{2},
 \and Alexander Morozov\aff{1}\corresp{\email{alexander.morozov@ed.ac.uk}}}

\affiliation{
\aff{1} School of Physics and Astronomy, The University of Edinburgh, James Clerk Maxwell Building, Peter Guthrie Tait Road, Edinburgh, EH9 3FD, United Kingdom
\aff{2} School of Mathematics and Maxwell Institute for Mathematical Sciences, University of Edinburgh, Edinburgh, EH9 3FD, United Kingdom
}

\begin{document}

\maketitle

\begin{abstract}
Successfully performing direct numerical simulations of polymeric flows remains a major challenge in computational fluid mechanics. In addition to the velocity field, such simulations must resolve polymeric degrees of freedom, often expressed via the conformation tensor, $ \c$, which captures the local stretch of polymer molecules. A key difficulty here lies in maintaining the physical requirement $\mathrm{Tr}\, \c>3$, which is not explicitly enforced by the governing equations. Consequently, simulations initiated from physical conditions may silently drift into unphysical states with $\mathrm{Tr}\, \c<0$, indicating a loss of positive-definiteness of the conformation tensor. Existing numerical methods to prevent this are costly, making direct numerical simulations of chaotic polymer flows, such as elastic turbulence, heavily reliant on high-performance computing.

Here, we ask whether simulations that violate $\mathrm{Tr}\, \c>3$ can still yield meaningful physical insight into the underlying dynamics. 
Using a standard pseudo-spectral method without explicit enforcement of positive-definiteness, we simulate a model dilute polymer solution driven through a plane channel at low Reynolds number and observe the transition to elastic turbulence. Our simulations exhibit two threshold resolutions: below the first, they become numerically unstable and exhibit a finite-time blow-up; above the second, they maintain positive-definiteness. In between, simulations remain stable and chaotic despite local violations of $\mathrm{Tr}\,  \c>3$. Surprisingly, these violations do not affect mid-plane statistics of velocity, its gradients, or polymer stretch, which match results from fully positive-definite simulations. This suggests that resolving flow structures or key flow statistics may not require the extreme resolutions needed to preserve positive-definiteness, potentially lowering computational barriers for studying elastic turbulence. We conclude by discussing strategies to leverage this insight for more accessible simulations.
\end{abstract}

\section{Introduction}
\label{sec:intro}

Understanding instabilities in flows of complex fluids is one of the outstanding problems of fluid mechanics \citep{Larson1999}. An archetypal example of such a challenge is posed by elastic (ET) \citep{Groisman2000,Steinberg2021} and elasto-inertial (EIT) \citep{Samanta2013,Dubief2023} turbulence -- chaotic flows of polymer solutions at low and moderate levels of fluid inertia. 

The dynamics of such flows are commonly characterised by two dimensionless parameters: The Weissenberg number, $\Wi$, defined as the product of a characteristic shear gradient magnitude and the polymer relaxation time, quantifies the ability of the flow to stretch polymer molecules, while the Reynolds number, $\Re$, measures the relative importance of inertial and viscous effects. In this framework, ET corresponds to the regime of \emph{large} $\Wi$ and \emph{small} $\Re$, while EIT arises when strong elastic stresses interact with finite inertia. 

Recently, significant progress was made in understanding the transition to EIT and ET \citep{Steinberg2021,Steinberg2022,Datta2022,Dubief2023,Morozov2025preprint}, particularly in flows in straight channels, where it has been discovered that such flows are organised around coherent structures, similar to their Newtonian counterparts \citep{Graham2021}. In contrast to Newtonian turbulence, however, viscoelastic coherent structures are driven by the anisotropic distribution of elastic stresses that are localised either at the walls of the domain (the wall mode) or at its midplane (the centre mode). The distinction between wall and centre modes, rooted in linear stability analysis results \citep{Garg2018,Chaudhary2019,Chaudhary2021,Khalid2021,Khalid2021a,Datta2022,Sanchez2022}, is also reflected in the nonlinear states that organise EIT and ET. In the EIT regime, a growing body of experimental, numerical, and theoretical work has connected the wall modes to coherent near-wall dynamics resembling self-sustained (viscoelastic) Tollmien--Schlichting waves \citep{Samanta2013,Dubief2013,Sid2018,Choueiri2018,Lopez2019,Shekar2019,Shekar2020,Choueiri2021,Shekar2021,Dubief2023}. 
At significantly lower values of $\Re$, however, polymeric flows have been shown to organise around travelling-wave solutions: \emph{arrowheads} in the EIT and \emph{narwhals} in the ET regimes, respectively.  Such solutions, localised around the centreline/centreplane of the flow geometry, were reported in Kolmogorov flow \citep{Berti2008,Berti2010,Nichols2025,Lewy2025}, pressure-driven channel flow \citep{Page2020,Morozov2022,Buza2022a,Lellep2023,Beneitez2024,Lellep2024,Lin2025,Morozov2025preprint}, flow through a cylinder array \citep{Zhu2024preprint,King2026preprint}, and curved pipe flow \citep{Lu2026preprint}. A unified pathway linking the coherent-structure dynamics of ET and EIT was proposed by \cite{FoggiRota2024} and \cite{Beneitez2024}. 

The ET and EIT phenomena are driven predominantly by the anisotropic distribution of the polymer stress, while the velocity field is mostly responding to its changes. While the latter can routinely be measured in experiments, there are currently no reliable techniques to measure time- and space-resolved profiles of the former. Under such circumstances, the only reliable source of information about the polymeric stress field can be obtained from direct numerical simulations (DNS), which have been crucial in uncovering the physical scenario presented above. Such simulations, however, are notoriously difficult to perform owning to a broad range of numerical instabilities and artefacts arising at sufficiently high values of $\Wi$, commonly referred to as the High-Weissenberg Number Problem \citep{Owens2002}. While the recent work by Kerswell and colleagues indicate that some of these instabilities might be caused by the presence of the polymer stress diffusion term in the constitutive equation \citep{Beneitez2023,Beneitez2024a,Couchman2024,Lewy2024,Beneitez2025}, the main origin of the High-Weissenberg Number Problem is rooted in the loss of positive-definiteness of the polymer conformation tensor $\c$.

There are two ways to introduce the polymer conformation tensor and understand the significance of its positive-definiteness. The first stems from kinetic theory of dilute polymer solutions \citep{Bird1987_2,Larson1999,Morozov2015}, which identifies $\c$ with the ensemble average dyadic product $\langle {\bm R}{\bm R}\rangle$, where ${\bm R}$ is the end-to-end vector of a polymer molecule. The conformation tensor is normalised with the typical size of the polymer molecule in thermal equilibrium such that in the absence of flow ${\c} = {\bm I}$, where ${\bm I}$ is the identity matrix. Depending on the microscopic force-extension law assumed, the polymer stress tensor ${\bm \tau}$ can then be expressed as an algebraic function of ${\c}$, that reduces to ${\bm \tau} \sim {\c} - {\bm I} = 0$, in equilibrium \citep{Alves2021}. Within this interpretation, under flow the eigenvalues of ${\c}$ can be viewed as the square of the polymer extension along its principle axes and should thus all be positive. Alternatively, one can introduce ${\c}$ without any appeal to kinetic theory. Within such an approach, the conformation tensor plays the role of an auxiliary variable related to the stress tensor through an algebraic relation. For a wide class of constitutive models, it is then possible to prove mathematically that the equation of motion for ${\c}$ is evolutionary: if at time $t_0$, the eigenvalues of ${\c}$ are non-negative everywhere in the domain, they remain non-negative at any later time $t>t_0$ \citep{Hulsen88,Hulsen90}. Since the rest state, ${\c} = {\bm I}$, is clearly non-negative definite, the eigenvalues of ${\c}$ should always remain non-negative. For a wide class of models, this bound can be made stronger, identifying physical solutions with the requirement that $\mathrm{Tr}\, {\c}>d$ everywhere in the domain, where $\mathrm{Tr}$ denotes the trace and $d$ is the number of spatial dimensions \citep{Hu2007,Yerasi2024}. In this work, we employ this stronger condition as an indicator of physical admissibility of a particular conformation tensor (and, hence, $\bm \tau$) distribution.

Unfortunately, this criterion is easily violated in DNS, even when staring from a positive-definite state, where accumulation of numerical errors leads to the loss of positive-definiteness and makes the results unphysical. There exist various numerical strategies of mitigating this problem, including changes of variables (e.g., the log-conformation \citep{Fattal2004}, square-root \citep{Balci2011}, higher-root \citep{Alves2021} or eigendecomposition \citep{Vaithianathan2003} formulations), adaptive diffusivity \citep{Dzanic2022, Dzanic2022c}, or intrinsically dissipative space discretisation \citep{Min2001,Kurganov2000,Dubief2005,Dubief2023}. Such strategies, however, are numerically costly and are often difficult to combine with highly-accurate spectral methods due to the infinite-order non-linearities involved in such strategies, thus leading to an infinite-order dealiasing problem. The demand for spatial accuracy often outweighs the concerns of preserving the positive-definiteness of the conformation tensor, and spectral simulations are usually performed in primitive variables thus allowing for potentially unphysical values of $\c$. While in some cases this results in a numerical instability and a finite-time blow-up, often such unphysical simulations exhibit superficially reasonable behaviour. Therefore, to ensure physically-meaningful results, each viscoelastic simulation has to be explicitly checked to preserve the positive-definite nature of the conformation tensor (or $\mathrm{Tr} {\c}>d$) everywhere in the domain at all times. 


To date, however, a significant fraction of publications reporting DNS results for ET or EIT do not explicitly comment on positive-definiteness of ${\c}$ and it is unclear whether the checks described above have been implemented as a routine part of the numerical protocol. With the currently growing number of DNS of chaotic flows of polymeric fluids, there is a possibility that some of these simulations do not satisfy the stringent criterion described above, and are thus formally unphysical. There is therefore an urgent need to access the importance of strictly preserving the positive-definite nature of the conformation tensor and to determine whether quantitatively correct results can still be obtained in formally unphysical simulations.

Here, to determine to what extent strict preservation of positive-definiteness is required for obtaining quantitatively reliable results, we perform DNS of a model polymeric fluid driven by an externally applied pressure gradient through a straight channel.
Based on our previous work, we start from the recently discovered three-dimensional narwhal states that organise ET in this system \citep{Lellep2024}, and construct a series of simulations at different resolutions designed to continuously switch from strictly positive-definite conformation tensor to the situations where $\mathrm{Tr} {\c}>d$ is strongly violated in scattered, highly localised regions of the domain.
In particular, these narrow regions are found at the interface between near-laminar regions and regions of large polymeric stress.
Surprisingly, we discover that beyond a certain numerical resolution required to ensure numerical stability, the presence of unphysical values of $\c$ does not alter the statistical properties of ET as quantified, for instance, by the velocity, velocity-gradient and polymer stretch distribution functions. On the basis of these results, we provide guidance for the cost-effective numerical simulation of elastic and elasto-inertial turbulence in wall-bounded shear flows.

\section{Governing equations and numerical details}

We consider a model polymer fluid confined within a straight, three-dimensional channel formed by the gap between two parallel, infinite plates. The geometry is conveniently described using a Cartesian coordinate system $(x,y,z)$, aligned with the streamwise, wall-normal, and spanwise directions, respectively. A constant external pressure gradient is imposed along the streamwise direction to drive the flow. The fluid motion is governed by the dimensionless simplified Phan-Thien Tanner (sPTT) constitutive model \citep{PhanThien1977} given by
\begin{align}
\label{eq:ptt} 
& \partial_t \c
+ {\bm u}\cdot\nabla{\bm \c} - \left(\nabla {\bm u}\right)^T\cdot{\bm \c} - {\bm \c}\cdot\left(\nabla {\bm u}\right) = \kappa \nabla^2 {\bm \c} 
- \frac{{\bm \c}-\bm{I}}{\Wi}\Bigg[ 1 + \epsilon\, \mathrm{Tr}\big({ \c} - \bm{I}\big)\Bigg],  \\
& \label{eq:ns} 
\partial_t  {\bm u} + {\bm u}\cdot\nabla{\bm u}  =
 -\nabla p + \frac{\beta}{\Re} \nabla^2{\bm u} + \frac{1-\beta}{\Re\,\Wi}\nabla\cdot{\bm \c} + \frac{2}{\Re}\hat{\bf x}, \\
&\label{eq:incomp} 
\nabla\cdot {\bm u} = 0.
\end{align}
Here, $\bm \c$ is the polymer conformation tensor, $\bm u$ is the fluid velocity, $p$ is the pressure, and $\hat{\bf x}$ is a unit vector in the streamwise direction. The superscript $(\cdot)^T$ denotes the transpose and $\bm{I}$ is the identity matrix. The governing equations are rendered dimensionless by using the following characteristic scales: all lengths are scaled with the channel half-width $d$, velocities with $U_0$, time with $d/U_0$, stress with $\eta_p U_0/d$, and pressure with $(\eta_s+\eta_p)U_0/d$. Here, $\eta_s$ and $\eta_p$ are the solvent and polymeric contributions to the viscosity, respectively, while the velocity scale $U_0$ corresponds to the laminar centreline velocity of a Newtonian fluid with the viscosity $\eta_s+\eta_p$ at the same value of the applied pressure gradient.

The flow is controlled by several dimensionless quantities: the strength of fluid elasticity is described by the Weissenberg number, $\Wi=\lambda\,U_0 / d$, the relevance of inertia - by the Reynolds number, $\Re=\rho\,U_0 d/(\eta_s+\eta_p)$, 
the relative contribution of the polymers to the fluid viscosity - by the ratio $\beta = \eta_s/(\eta_s + \eta_p)$, while the degree of shear-thinning is set by $\epsilon$. Here, $\rho$ is the density of the fluid, $\lambda$ is its Maxwell relaxation time, and $D$ is the diffusion coefficient of a polymer molecule. The above equations include a (global) polymer diffusion term where the stress diffusivity $\kappa = D / d\,U_0$ compares the typical distance diffused by a polymer molecule in one time unit with the channel half-width. In what follows, time intervals are reported in terms of $t/\Wi$, which in our dimensionless units corresponds to measuring physical time in terms of the Maxwell relaxation time $\lambda$. 

To simulate an unbounded flow in the streamwise and spanwise directions, these directions are set to be periodic, with the domain lengths $L_x = L_z = 10$. At the walls, we impose no-slip boundary conditions for the velocity, i.e. ${\bm u}(x,y=\pm 1,z,t)=0$,  while the boundary conditions for the conformation tensor are obtained by requiring that ${\c}$ at the walls is equal to the corresponding value obtained by solving the constitutive model Eq.~\eqref{eq:ptt} with $\kappa=0$ \citep{Thomas2006}.

\begin{table}
\begin{center}
\captionsetup{width=\columnwidth}
\def~{\hphantom{0}}
\begin{tabular}{ccccccc}
\hline
\hline
& $(N_x, N_y, N_z)$ & $M_{dof}/M_{dof}^{(A1)} $ & ${\chi}/10^{-4}$ & $\mathcal{P}{\left(\mathrm{Tr}\c<3\right)}/10^{-5}$  & $T/\Wi$\\
\hline
A1 & $(128, 512, 128)$ & 1 & 7.2 & 6.6  & 68.0\\
            
A2 & $(512, 128, 128)$ &  1 & 3.1 & 2.9 & 15.0\\

A3 & $(256, 256, 256)$ & 2  & 1.2 & 1.1 & 15.0\\

A4 & $(512, 256, 128)$ &  2 & 2.9 & 2.8 & 15.0\\

A5 & $(512, 128, 256)$ &  2 & $1.1 \! \times \! 10^{-3}$ & $1.0 \! \times \! 10^{-3}$ & 15.0 \\

A6 & $(512, 128, 512)$ &  4 & 0 & 0 & 68.0\\

\hline
\hline
\end{tabular} 
\caption{ 
Summary of the simulations used in this work. Here, $(N_x, N_y, N_z)$ describes the spatial resolution, $\chi$ is the fraction of midplane points where $\mathrm{Tr}\c<3$, $\mathcal{P}{\left(\mathrm{Tr}\c<3\right)}$ is the probability of such events (see eqs.~\eqref{eq:prob_neg}-\eqref{eq:chi}), while $T/\Wi$ is the total run time in polymer relaxation time units. For all datasets, snapshots have been sampled each $0.67 \,T/\Wi$ time units. The total number of degrees of freedom in a simulation, $M_{dof} = 10\, N_x N_y N_z$, corresponding to the spectral decomposition of 6 independent components of $\bm c$, three components of $\bm u$, and one component of $p$, is compared to that of the A1 run, $M_{dof}^{(A1)} \sim 8.4 \times 10^7$.}
\label{tab:datasets}
\end{center}
\end{table}  

The equations of motion, Eqs.~\eqref{eq:ptt}-\eqref{eq:incomp}, are solved numerically using an MPI-parallel fully-dealiased pseudo-spectral code developed within the Dedalus framework \citep{Burns2020}. Time-discretisation is performed using a 4th-order semi-implicit BDF scheme \citep{Wang2008} with the timestep $2.5 \times 10^{-3}$.
The velocity, conformation tensor, and pressure fields are represented through a spectral decomposition based on Fourier-Chebyshev-Fourier modes in the streamwise, wall-normal, and spanwise directions, respectively. The spectral resolution is set by specifying the number of modes, $(N_x,N_y,N_z)$, used in each direction. 

In this work, we focus on a single set of fluid parameters given by $\Wi=150$, $\Re=10^{-2}$, $\beta=0.9$, $\epsilon=5\times 10^{-5}$, and $\kappa=5\times 10^{-5}$. 
Compared to the parameters used in \cite{Lellep2024}, the current fluid is significantly more elastic and less shear thinning, intentionally making it more prone to developing numerical instabilities and making it more difficult to ensure positive-definiteness of the conformation tensor. 
We assess the influence of the spatial resolution on the positive-definiteness of the conformation tensor by performing a set of numerical simulations with systematically increased $N_x$, $N_y$, and $N_z$. This process was terminated when we reached the resolution that strictly preserved the PD nature of $\c$ at all times. The details of all simulations are given in Table \ref{tab:datasets}.  We note that all simulations run at resolutions below that of A1 exhibited a finite-time blow-up and were therefore excluded from the analysis.

To ensure meaningful comparison, all simulations presented below were started from the same initial condition. To this end, we performed a long simulation (not shown) for the current set of parameters with $N_x=N_z=256$ and $N_y=1024$, until it reached statistically-steady elastic turbulence, following the protocol from \cite{Lellep2024}. We then selected a single, positive-definite configuration from the statistically steady-state of that run and re-interpolated it on the A1-A6 spatial grids (see Table \ref{tab:datasets}). As confirmed below in Fig.~\ref{fig:min_trC}, all initial conditions thus obtained are positive-definite, even at the lowest resolution considered.


\section{Results}

\begin{figure}
\centering
\captionsetup{width=\columnwidth}
\includegraphics[width=0.8\columnwidth]{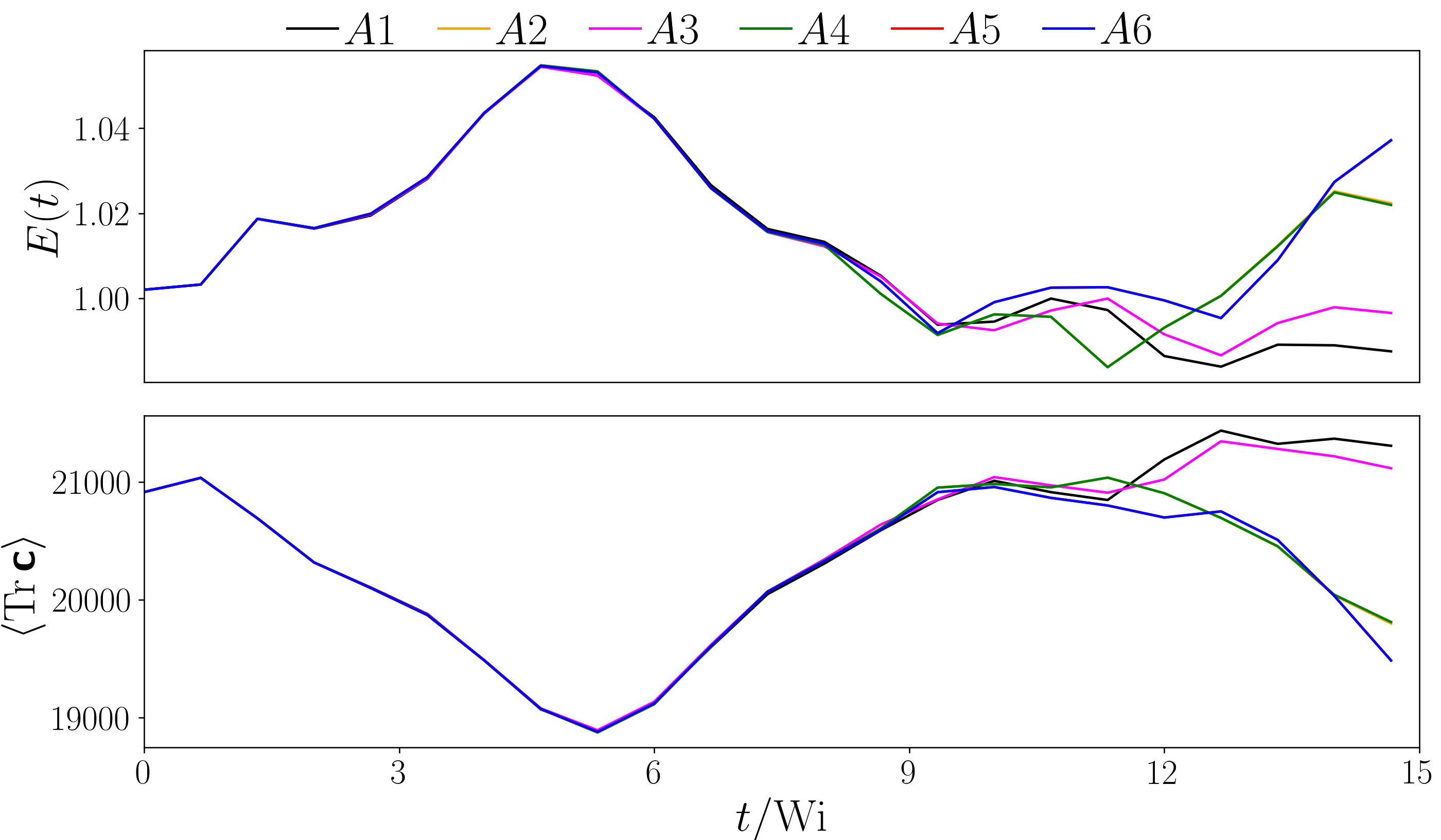}
\caption{
Short-time evolution of the total kinetic energy $E(t)$ (top panel) and the average polymer stretch $\langle\mathrm{Tr}\, { \c}\rangle$ for all datasets.}
\label{fig:timeseries}
\end{figure}

Here, we present the results of the A1-A6 runs started from the same initial condition, as discussed above. Since we are dealing with a chaotic system, small numerical differences between the initial conditions at different resolutions will amplify in the course of the simulations, and we expect the associated chaotic trajectories to lose resemblance at sufficiently long times. To quantify this behaviour, we introduce two observables, the kinetic energy $E(t) = \langle ||{\bm u}({\bm r},t)||^2/2 \rangle$, and a measure of the total polymer stretch $\langle\mathrm{Tr}\, {\bm \c}({\bm r},t)\rangle$, where $\langle \dots \rangle$ denotes the volume average and ${\bm r}$ is the spatial position in the simulation domain. In Fig.~\ref{fig:timeseries}, we monitor the time-evolution of these quantities and confirm that the trajectories follow each other closely for $t/\Wi\lesssim 8$. This timescale is in line with the typical stress correlation times observed previously by \cite{Lellep2024} in elastic turbulence at somewhat different fluid parameters.

\begin{figure}
\centering
\captionsetup{width=\columnwidth}
\includegraphics[width=0.81\columnwidth]{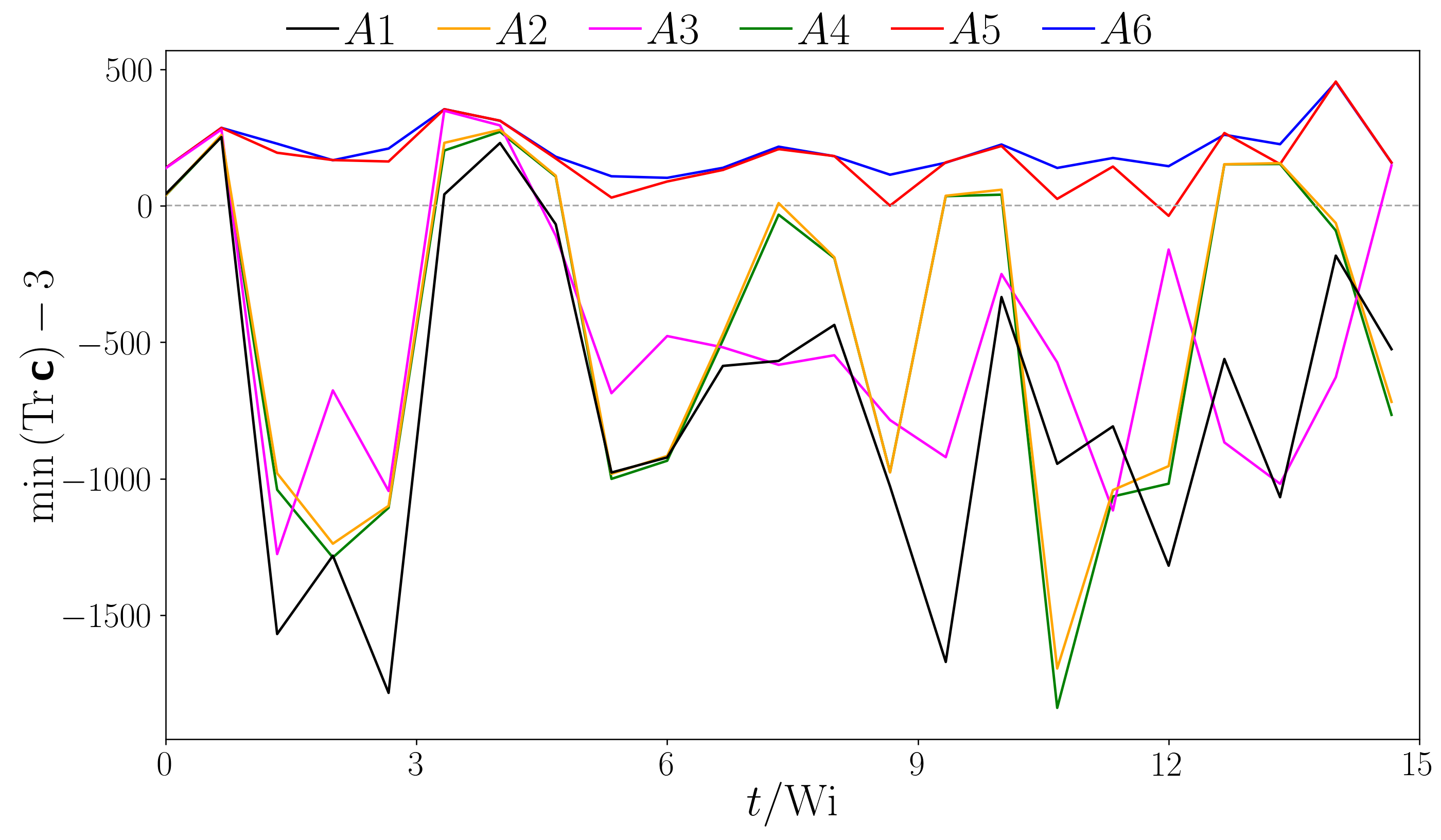}
\caption{
Instantaneous value of $\min(\mathrm{Tr}\, \c - 3)$ as a function of time. The dashed line indicates the admissibility limit, $\min(\mathrm{Tr}\, \c) = 3$.
}
\label{fig:min_trC}
\end{figure}

First, we focus on the short-time behaviour of the A1-A6 simulations. Although Fig.\ref{fig:timeseries} suggests that the global scalar quantities, such as the kinetic energy and the polymer stretch, are virtually indistinguishable between different runs, the PD properties of the corresponding conformation tensors vary significantly. This observation is quantified in Fig.\ref{fig:min_trC}, where we plot the global minimum of $\mathrm{Tr}\, \c$ within the simulation domain as a function of time. Since a physical conformation tensor requires $\mathrm{Tr}\, \c>3$ at each point in space \citep{Hu2007,Musacchio2003}, observing a negative value of $\mathrm{Tr}\, \c -3$ implies 
the presence of numerical artefacts and unphysical values of $\c$.
As can be seen from Fig.\ref{fig:min_trC}, the loss of positive-definiteness occurs for most of the resolutions, with only the A6 simulation remaining strictly positive-definite at all times. While the minimum value of $\mathrm{Tr}\, \c$ seems to be rather insensitive to the resolution in the wall-normal direction, $N_y$, sufficient numbers of Fourier modes $N_x$ and $N_z$ are required to ensure $\mathrm{Tr}\, \c \geq 3$.

\begin{figure}
\centering
\captionsetup{width=\columnwidth}
\includegraphics[width=0.95\columnwidth]{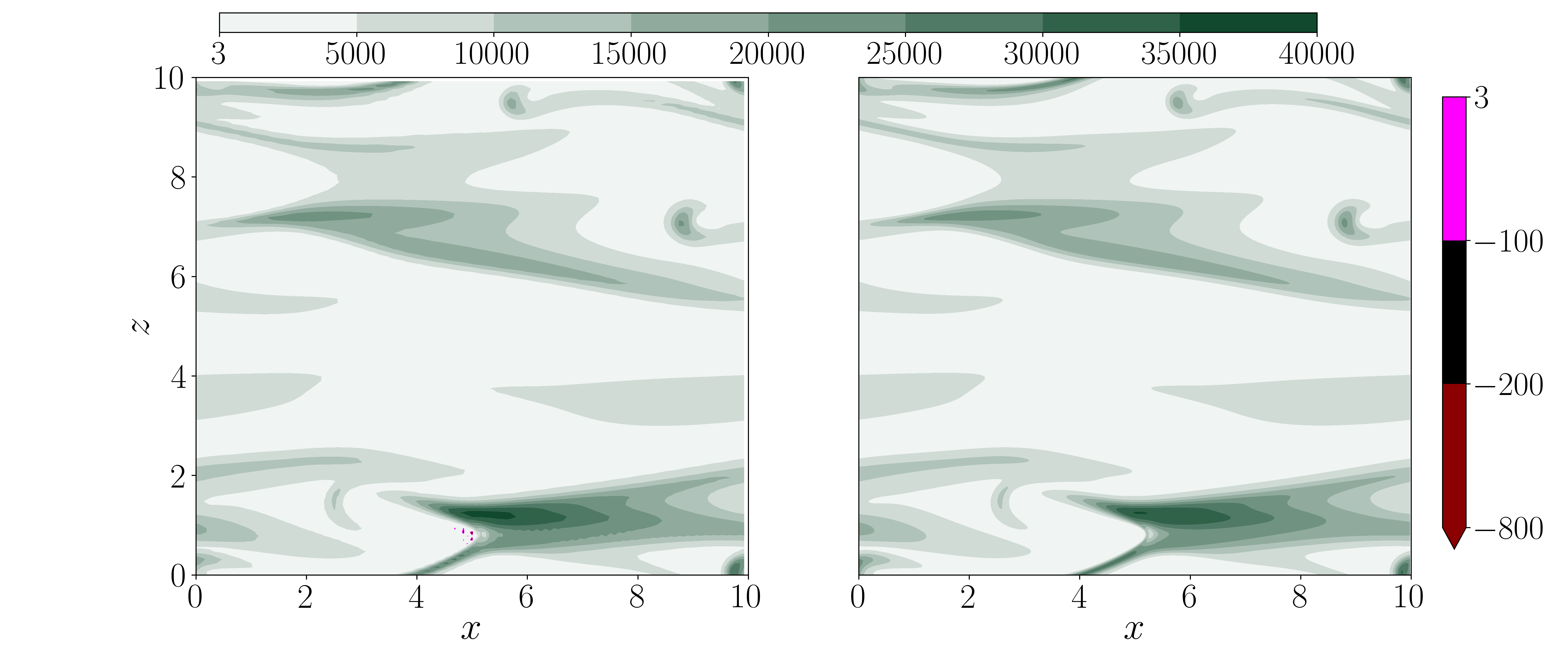}
\makebox[\textwidth]{%
\hspace{-0.2cm}
\includegraphics[width=0.62\columnwidth]{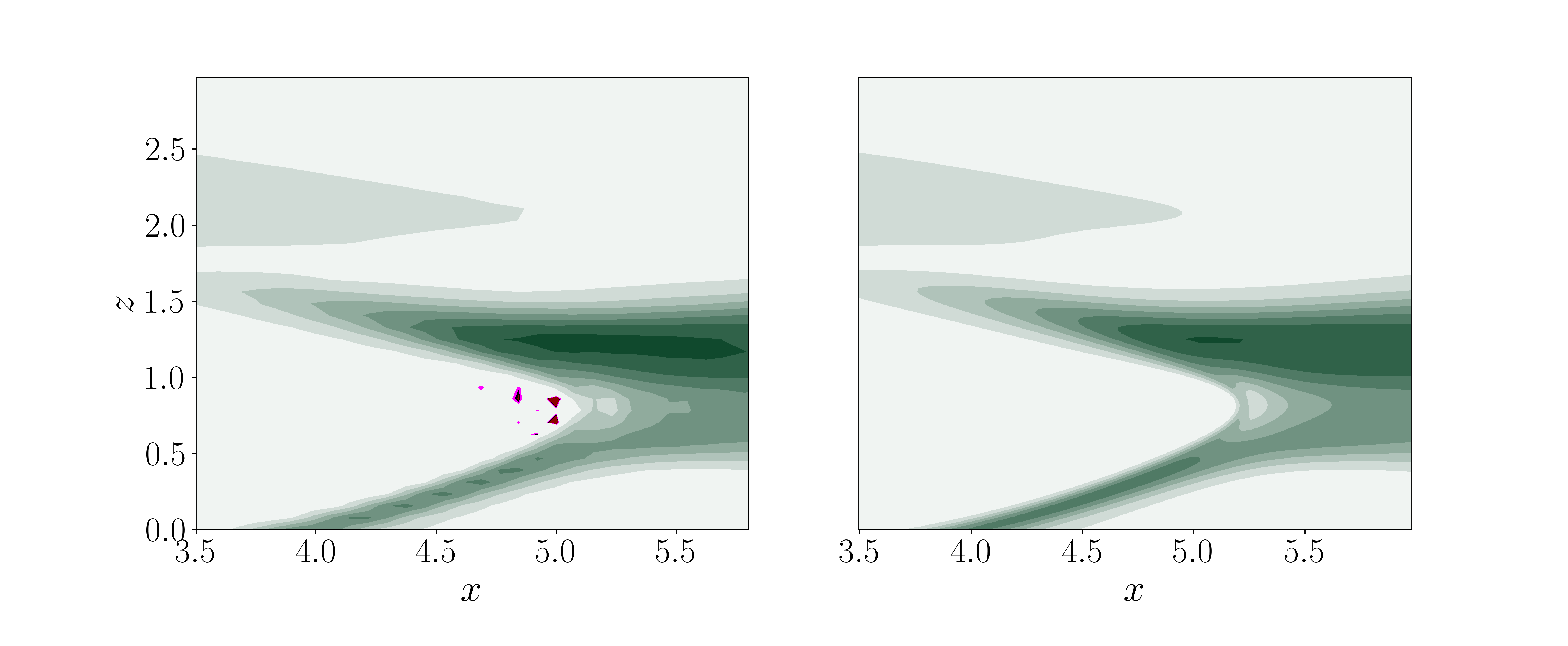}}
\caption{
Midplane profiles of $\mathrm{Tr}\,\c$ at $t/\Wi=2.7$ from the A1 (left panels) and A6 (right panels) runs, respectively. The bottom-row panels are magnifications of the top-row images for $3.5 \leqslant x \leqslant 5$ and $0 \leqslant z \leqslant 3$. Physically inadmissible points with $\mathrm{Tr}\,\c < 3$ can be seen close to the bottom-center coherent structure in the left panels. 
}
\label{fig:pd_vis}
\end{figure}

\begin{figure}
\centering
\captionsetup{width=\columnwidth}
\includegraphics[width=0.8\columnwidth]{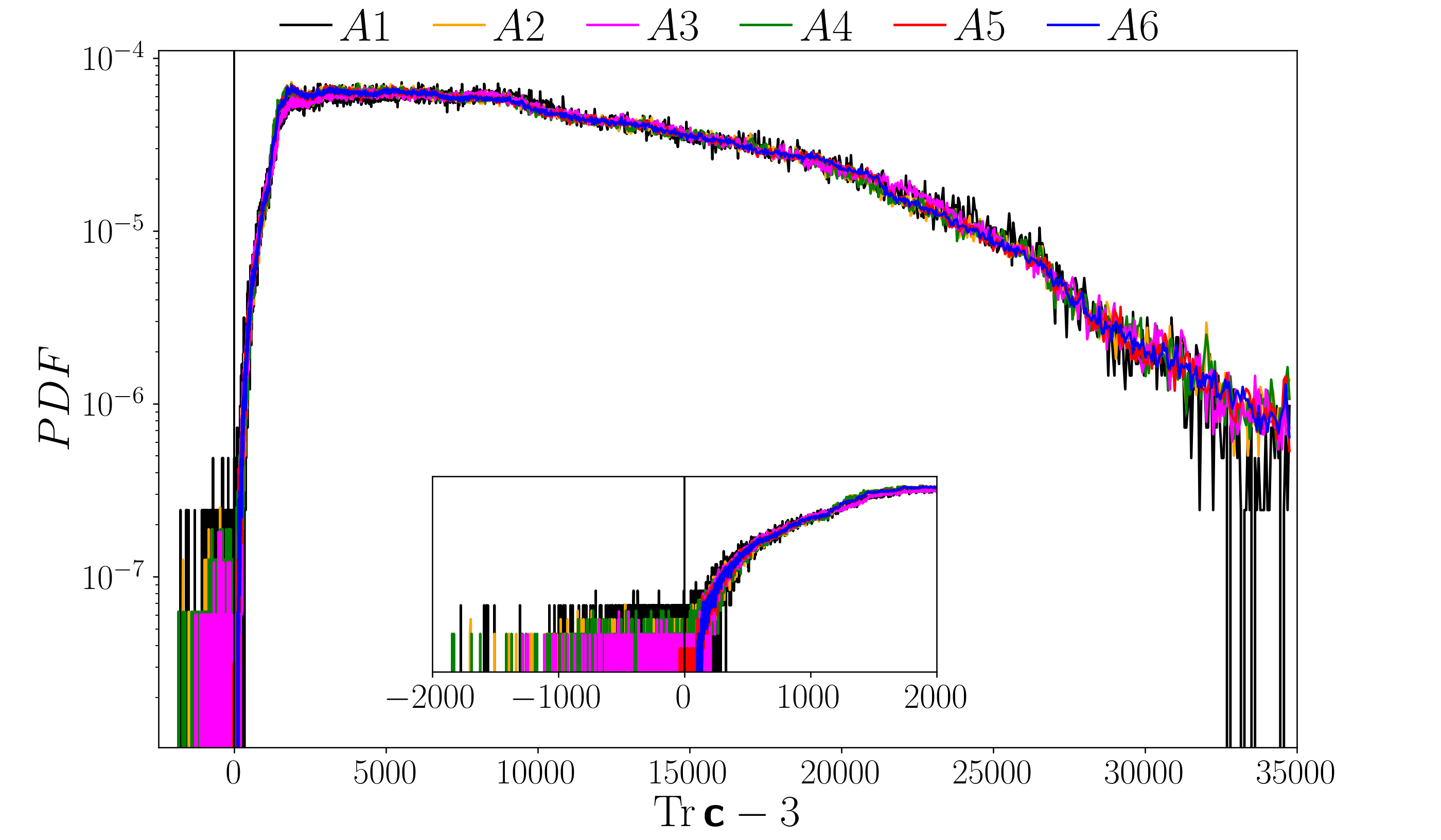}
\caption{
PDF of $\rm Tr \, \mathsfbi{c}$ at the midplane. The sampling time is that of the data shown in Figs.~\ref{fig:timeseries} and \ref{fig:min_trC}. The inset is a magnification of the $x$-axis region $[-2000, 2000]$. The vertical dark grey line indicates $\rm Tr \, \mathsfbi{c} = 3$.
}
\label{fig:pd_pdfs_trC_test}
\end{figure}

In addition to the degree to which the PD properties of $\c$ are violated, we quantify how often and where they are violated for a given resolution. As was noted in the Introduction, elastic turbulence in channel flows is characterised by spot-like coherent structures strongly localised around the channel midplane \citep{Lellep2024}. Since the largest deviations of the elastic stress from its laminar value take place around the midplane, below we focus on the analysis of the polymer stretch at the midplane, $\mathrm{Tr}\, \c (x,y=0,z,t)$. We confirmed that all violations of positive-definiteness in our simulations happen in a narrow vicinity of the midplane 
plane. In Fig.~\ref{fig:pd_vis} we present the comparison of the polymer stretch 
for the A1 and A6 runs corresponding to the lowest and highest resolutions studied, respectively. We can see that unphysical stresses emerge in the vicinity of the stagnation regions of the coherent structures where the polymer stress is largest. To analyse the statistics of the midplane polymer stretch, in Fig.\ref{fig:pd_pdfs_trC_test} we present the PDF of $\mathrm{Tr}\, \c(x,y=0,z,t)$ constructed from the values of the midplane polymer stretch for all configurations recorded up to $t/\Wi\lesssim 8$. It allows us to make two important observations. Firstly, we note that the lower is the resolution, the larger is the number of spatial points where $\c$ loses its positive-definiteness.
We quantify this by introducing the probability of observing a local non-physical value of $\c$, 
\begin{equation}
\mathcal{P}{\left(\mathrm{Tr}\c<3\right)} = \int_{-\infty}^3 \mathrm{d} X\, P_c(X), 
\label{eq:prob_neg}
\end{equation}
where $P_c$ denotes the probability 
density function (PDF) of $\mathrm{Tr}\,  \c(x,y=0,z,t)$,
and the fraction of midplane points where such events occur,
\begin{equation}
    \chi = \dfrac{N_{events} \, \mathcal{P}\left(\mathrm{Tr}\c<3\right)}{N_x N_z}.
     \label{eq:chi}
\end{equation}
The denominator corresponds to the total number of degrees of freedom in the channel midplane, and $N_{events}$ is the total number of local measurements used to construct the PDF.
The short-time estimates for these quantities are reported in Table \ref{tab:datasets} for each dataset. As we can see, both $\mathcal{P}{\left(\mathrm{Tr}\c<3\right)}$ and $\chi$ systematically decrease with an increased resolution and can be made arbitrarily small by adjusting $N_x$, $N_y$, and $N_z$. 
The second key 
observation to be made from the data shown in Fig.~\ref{fig:pd_pdfs_trC_test} is that the PDFs of the physical values, $\mathrm{Tr}\, \c(x,y=0,z,t)>3$, are virtually indistinguishable and do not change with the resolution. 
This observation implies that the presence of unphysical events does not alter the global statistics of the polymer stress. 

\begin{figure}
\centering
\captionsetup{width=\columnwidth}
\includegraphics[width=0.9\columnwidth]{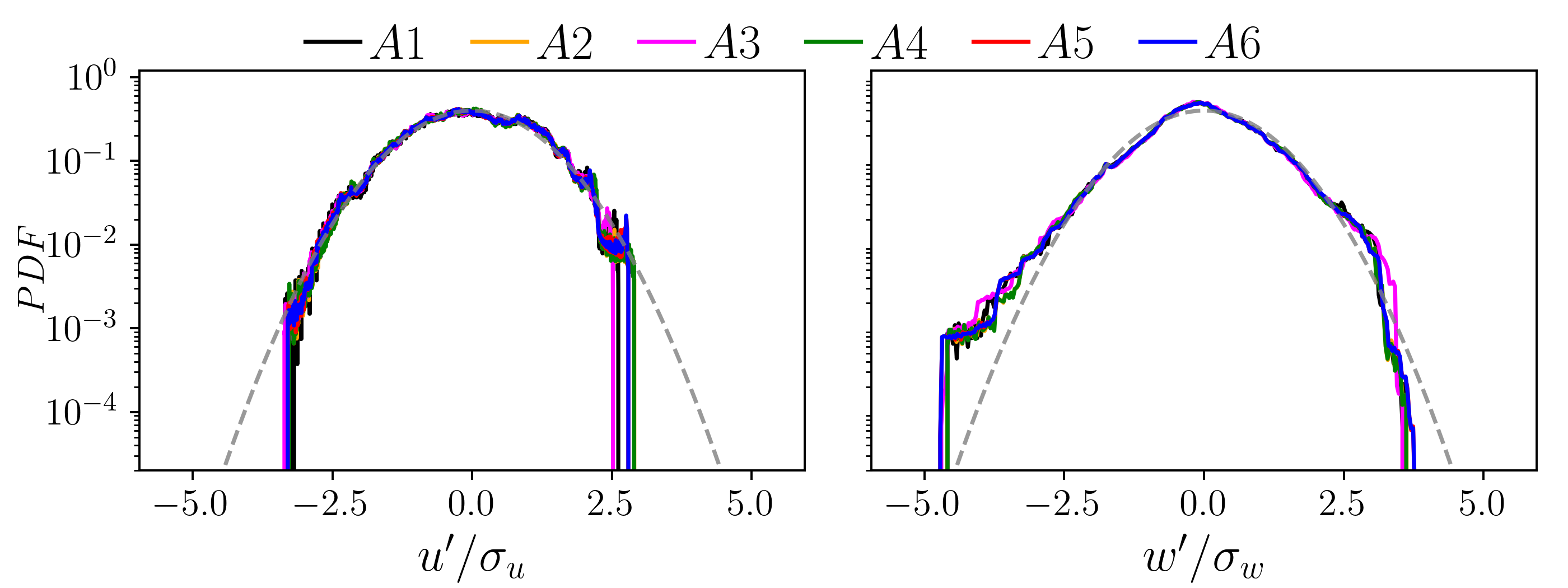}
\caption{
Standard PDFs of the streamwise (left) and spanwise (right) velocity component fluctuations for all the datasets, sampled at the midplane for $t/\Wi \in [0.0,9.2]$, with the standard deviations $\sigma_u$ and $\sigma_w$, respectively. The grey dashed line corresponds to a Gaussian with a zero mean and unit variance.
}
\label{fig:pd_pdfs_fields}
\end{figure}

\begin{figure}
\centering
\captionsetup{width=\columnwidth}
\includegraphics[width=0.9\columnwidth]{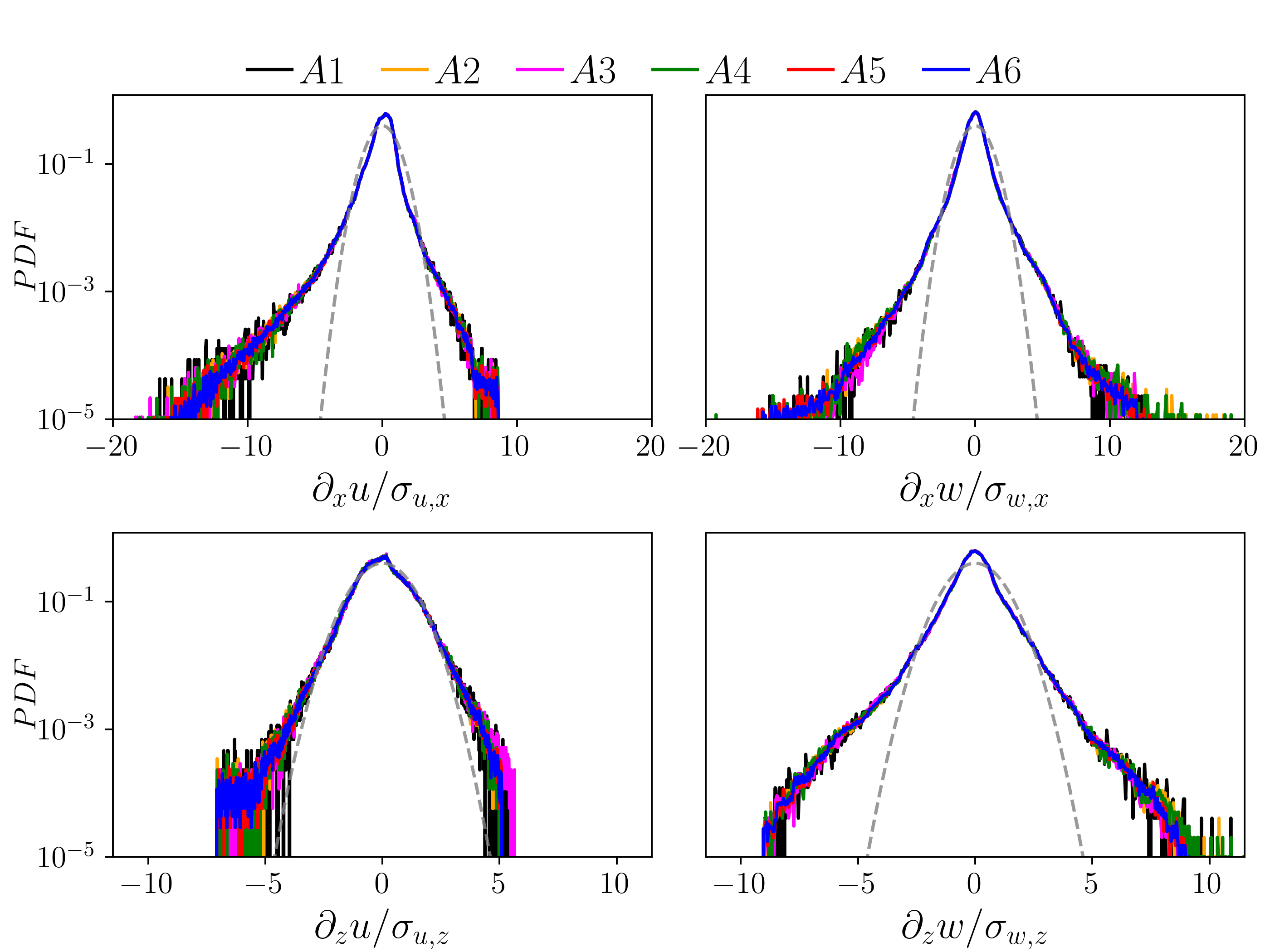} 
\caption{
Standard PDFs of longitudinal and transverse gradients of the streamwise (left) and spanwise (right) velocity components for all the datasets, sampled at the midplane for $t/\Wi \in [0.0,9.2]$, with the standard deviations $\sigma_{u,x}$, $\sigma_{u,z}$, $\sigma_{w,x}$ and $\sigma_{w,z}$, respectively. 
The grey dashed line corresponds to a Gaussian with a zero mean and unit variance.
}
\label{fig:pd_pdfs_grad}
\end{figure}

A key point to assess is the extent to which negative definite stresses alter flow structures and statistics at large and small scale. For a quantitative assessment, we therefore consider midplane velocity-field and velocity-gradient statistics. Fig.\ref{fig:pd_pdfs_fields}  presents the standardised PDFs of the streamwise and spanwise velocity fluctuations, $u'$ and $w'$, for all datasets. Most importantly, 
no significant difference can be observed in the velocity-components statistics across the datasets. The streamwise velocity fluctuations are approximately Gaussian irrespective of the resolution and the presence of negative definite stresses in datasets A1-A5. In comparison, extreme fluctuations of spanwise velocity are more likely, as can be seen from the super-Gaussian, exponential tails of PDFs, which again coincide across all datasets, except for a small amount of noise in the tails.
Fig.\ref{fig:pd_pdfs_grad} shows the PDFs of the longitudinal and transverse gradients of the streamwise and spanwise velocity components.
Similar to the velocity statistics, apart from the expected noise in the tails, we do not observe significant differences between the datasets.
That is, both velocity and velocity-gradient statistics are unaffected by the negative definite stresses present in datasets A1-A5.
In summary, coarse resolution that leads to negative-definite stresses does not alter the flow and stress statistics.

\begin{figure}
\centering
\captionsetup{width=\columnwidth}
\includegraphics[width=0.8\columnwidth]{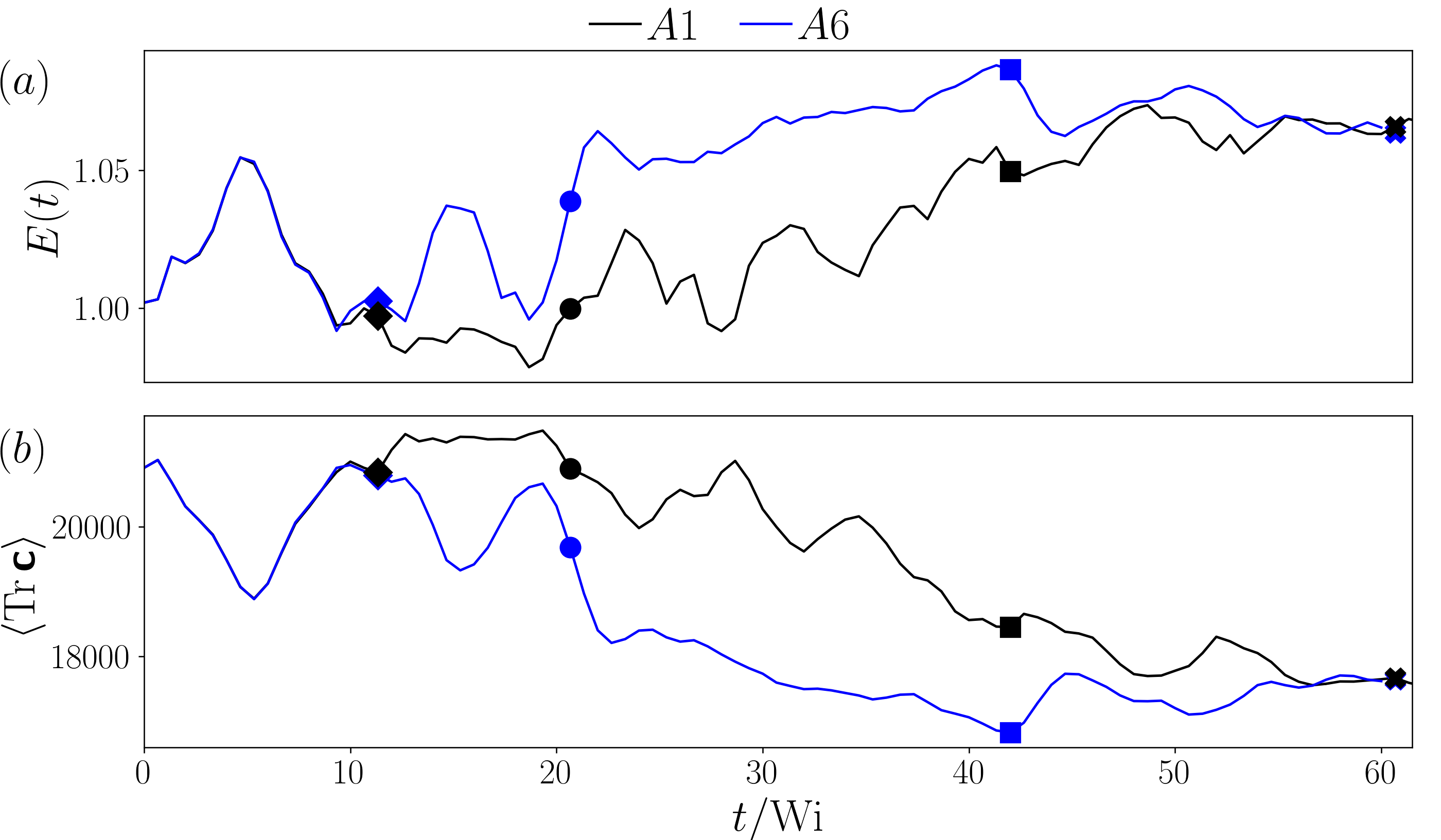}

\includegraphics[width=0.8\linewidth]{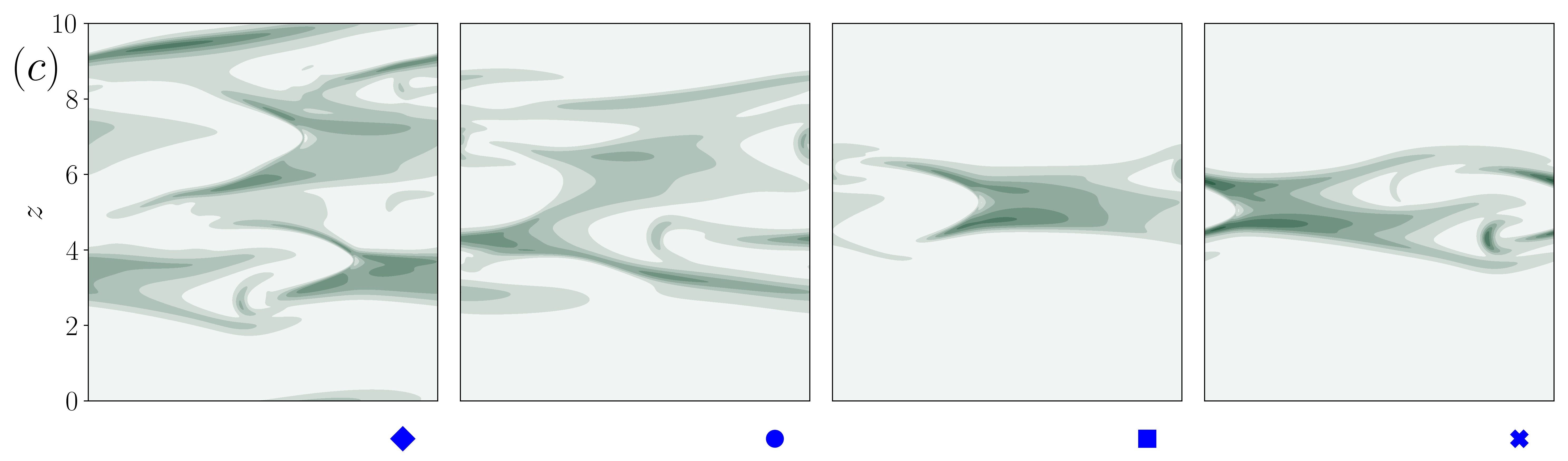} 

\hspace{0.07cm}  \includegraphics[width=0.8\linewidth]{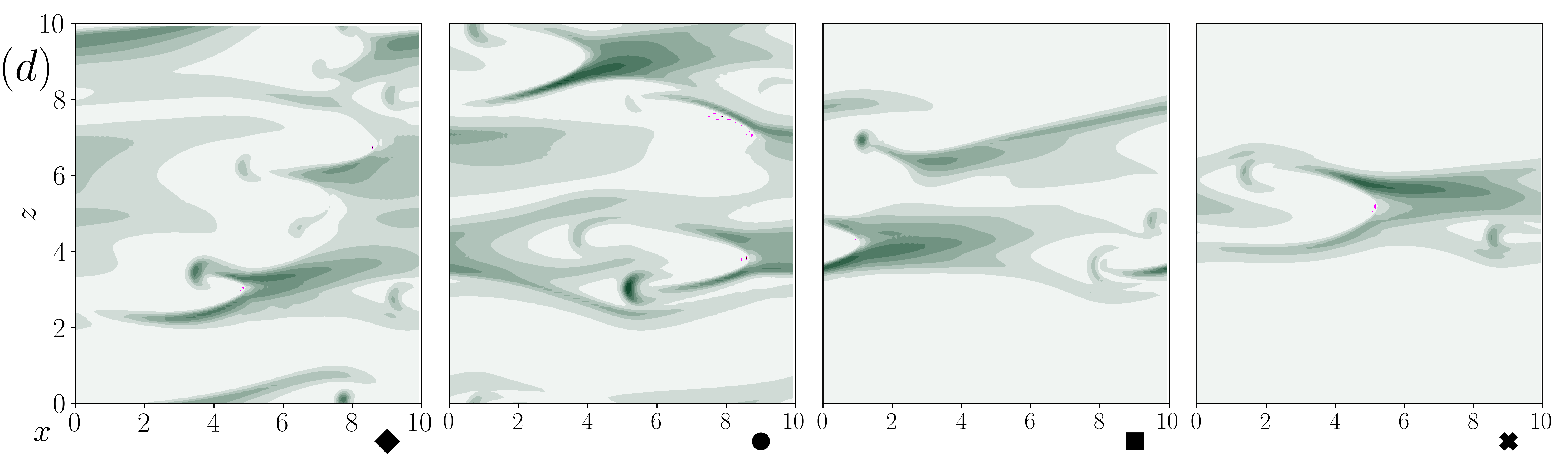} 
\caption{
(a,b): Long time series of (a) mean kinetic energy and (b) polymer extension for datasets A6 and A1. The symbols indicate instances in time corresponding to the midplane visualisations of $\rm{Tr} \, \mathsfbi{c}$ for A6 and A1 shown in (c) and (d), respectively. The colour scheme and extent are the same as in Fig.~\ref{fig:pd_vis}. Note the barely visible regions with $\rm{Tr} \, \mathsfbi{c} < 3$ in (d).
}
\label{fig:timeseries_single}
\end{figure}

%
The conclusion reached so far is based on short-time simulations that, by construction, visually appeared indistinguishable. We now demonstrate that the same conclusion holds in general, even for chaotic trajectories that do not resemble each other. To this end, we focus on the lowest, A1, and highest, A6, resolutions studied here, and continue the corresponding simulations to significantly larger values of $t/\Wi$. 
In Fig.\ref{fig:timeseries_single} we present the time-evolution of the kinetic energy and the polymer stretch, where the A1 and A6 trajectories strongly deviate from each other and 
thus are likely to have decorrelated for $t/\Wi\gtrsim 8$. As previously reported by \cite{Lellep2024}, elastic turbulence in channel flows exhibits spatio-temporal intermittency at sufficiently high $\Wi$, with the number of localised turbulent `spots' fluctuating strongly in time. This is indeed what we observe here as well. While both trajectories start from a configuration where a large portion of the midplane is occupied by turbulent spots, as can be seen from Fig.~\ref{fig:pd_vis}, at later times the number of spots decreases in both simulations, albeit at different $t/\Wi$, see Fig.~\ref{fig:timeseries_single}. 

In the following, we focus our analysis on the time interval $t/\Wi \geq 46$. For these times, both trajectories exhibit the same dynamical state with a single localised structure in the midplane, as can be seen in Fig.~\ref{fig:timeseries_single}, yet their spatial realisations are very different between the two runs. We repeat the analysis performed above for the early-time trajectories and demonstrate that the A1 and A6 runs are sampling two very similar statistical states despite the latter simulation having four times the number of the degrees of freedom of the former. This difference in the resolution becomes even more substantial along the $x$-$z$ midplane where A6 is 16 times more resolved than A1.  
\begin{figure}
\centering
\captionsetup{width=\columnwidth}
\includegraphics[width=0.8\linewidth]{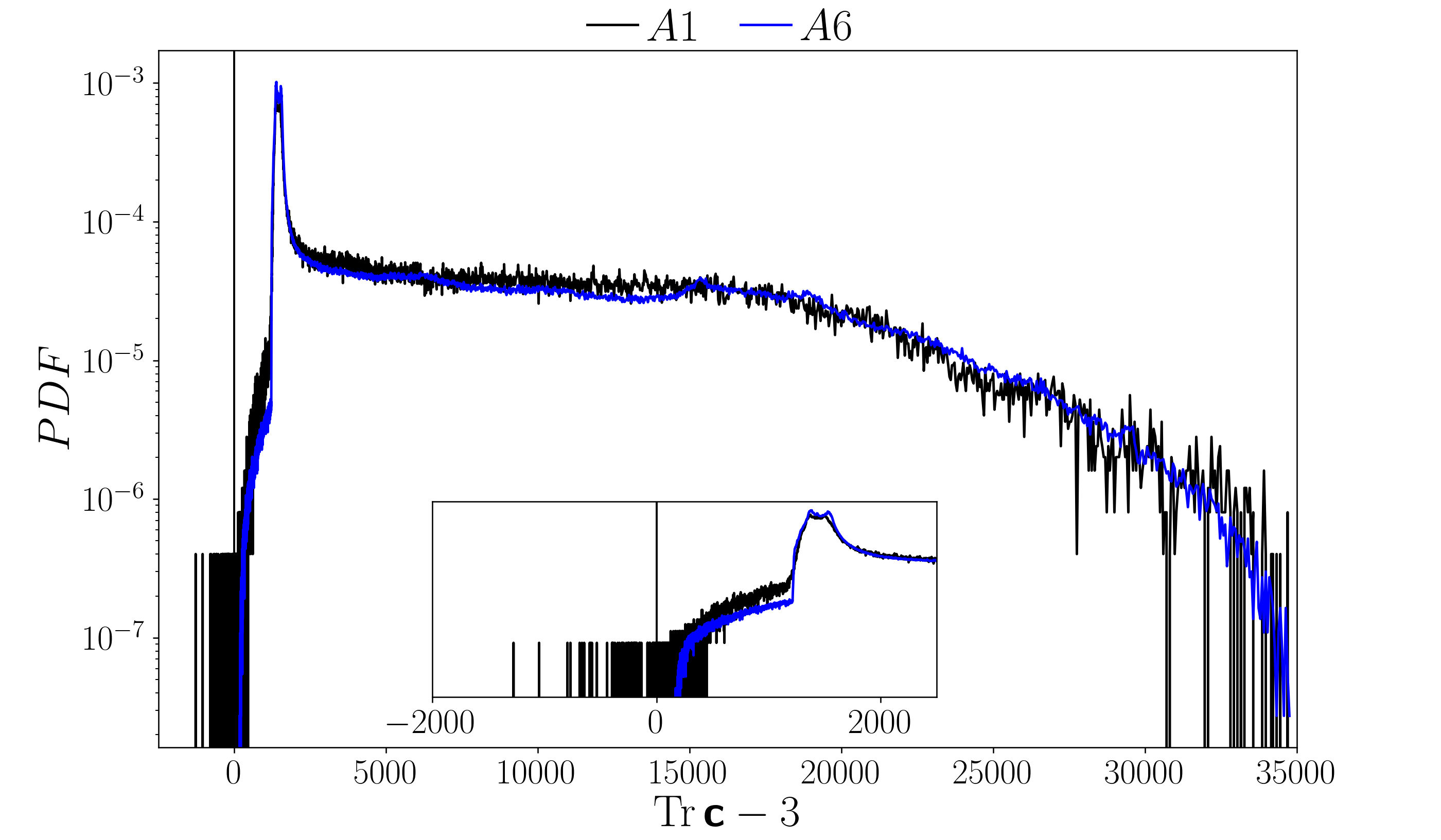}
\caption{
PDFs of $\rm Tr \, \mathsfbi{c} -3$ at the midplane for A1 and A6 in the single-structure state, sampled for $ 46 \leqslant t/Wi \leqslant 60$. The vertical dark grey line indicates $\rm Tr \, \mathsfbi{c} = 3$. The inset shows the PDF for values in an interval around zero.
}
\label{fig:trc_pdf}
\end{figure}

In Fig.~\ref{fig:trc_pdf} we present the PDF of the polymer stretch, similar to Fig.~\ref{fig:pd_pdfs_trC_test}, and observe that while the A1 run occasionally exhibits non-physical values at high-stress regions of the channel midplane (see Fig.\ref{fig:timeseries_single}), the A6 run remains strictly positive-definite at all times. For the physical values of $\mathrm{Tr}\, \c$, both PDF are virtually the same, sharing a peak located at $\mathrm{Tr}\, \c \approx 1500$ which is absent from the short-time PDF, see Fig.~\ref{fig:pd_pdfs_trC_test}. Since the latter is associated with a large number of the midplane coherent structures, we conclude that the presence of such a peak is a signature of a single turbulent spot.  

\begin{figure}
\centering
\captionsetup{width=\columnwidth}
\includegraphics[width=0.8\linewidth]{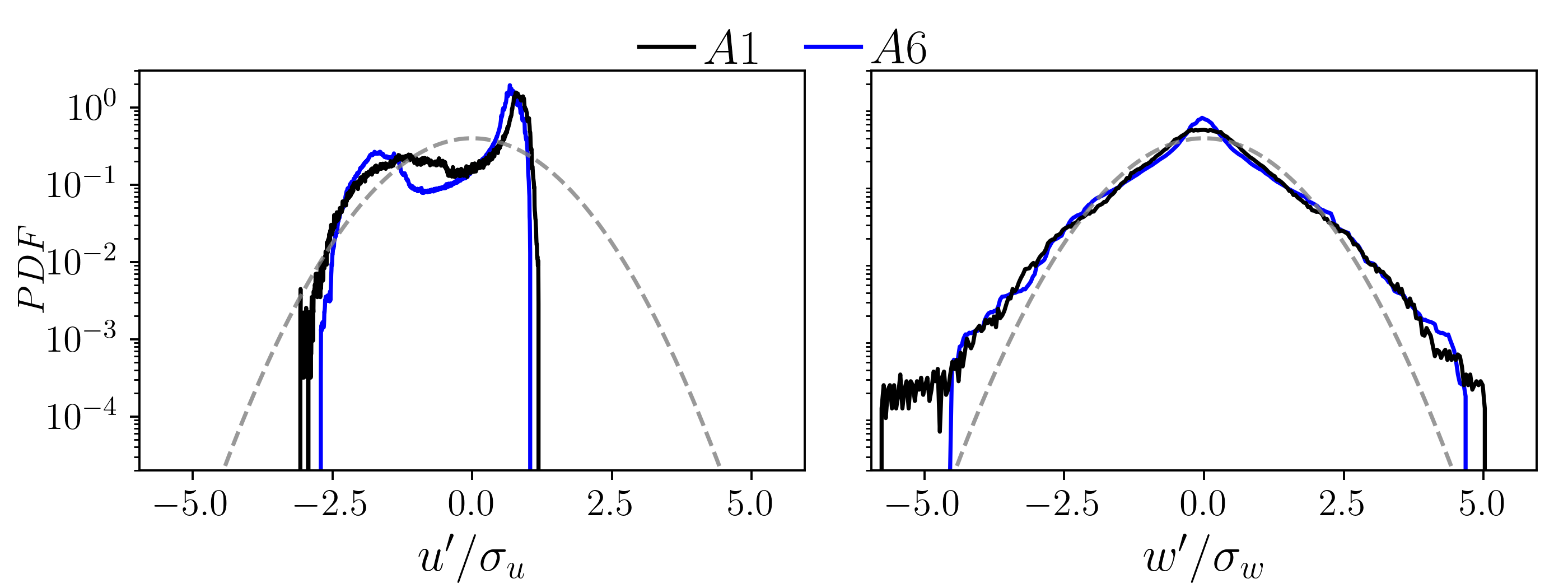}
\caption{
Standard PDFs of the streamwise (left) and spanwise (right) velocity component fluctuations for A1 and A6 in the single-structure state, with the standard deviations $\sigma_u$ and $\sigma_w$, respectively.  The sampling time corresponds to that of Fig.~\ref{fig:trc_pdf}.
The grey dashed line corresponds to a Gaussian with a zero mean and unit variance.
}
\label{fig:pdf_fields_single}
\end{figure}

\begin{figure}
\centering
\captionsetup{width=\columnwidth}
\includegraphics[width=0.8\linewidth]{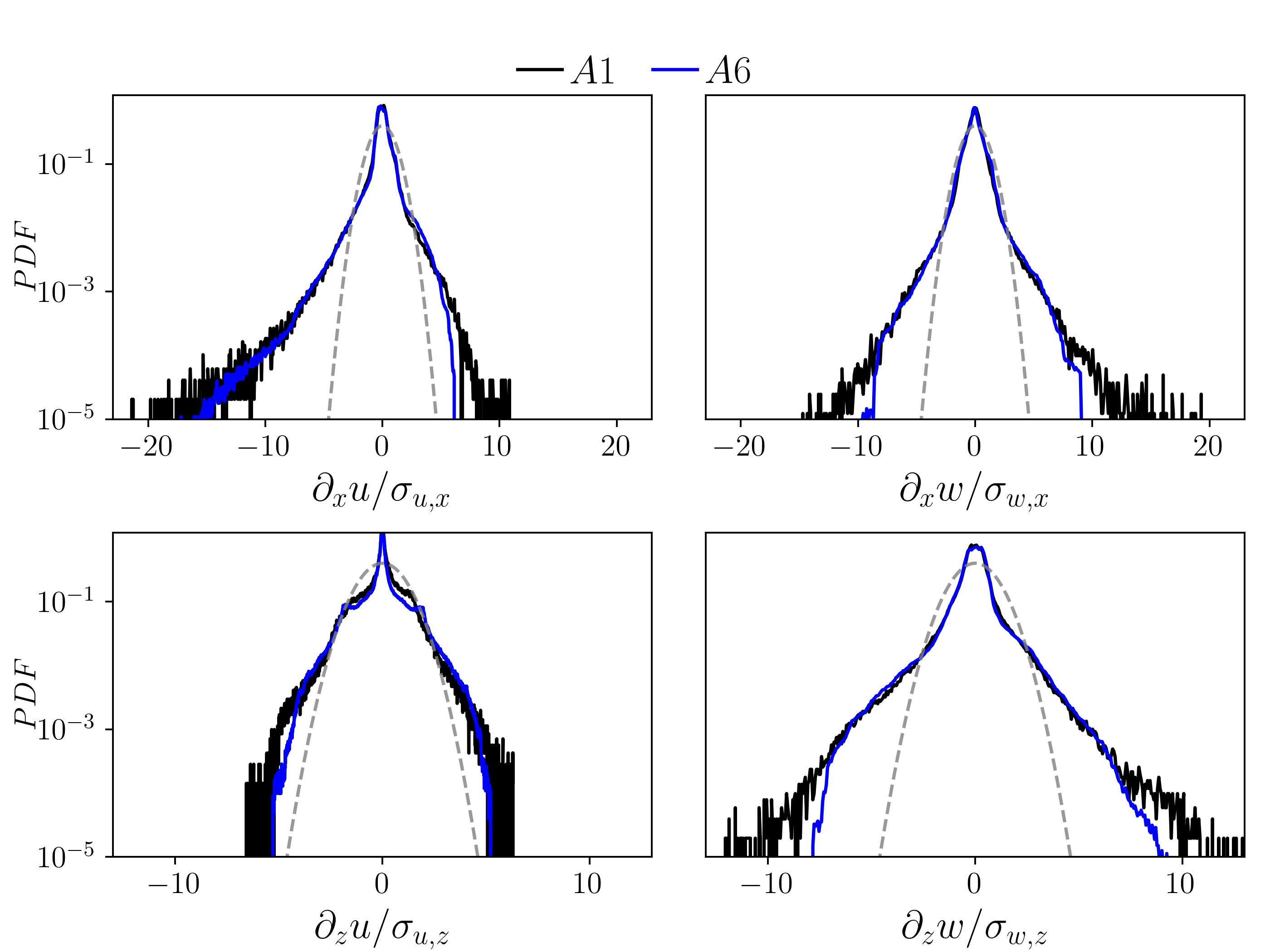}
\caption{Standard PDFs of the longitudinal and transverse gradients of the streamwise (left) and spanwise (right) velocity components for A1 and A6 in the single-structure state, with the standard deviations $\sigma_{u,x}$, $\sigma_{u,z}$, $\sigma_{w,x}$ and $\sigma_{w,z}$, respectively.  
The sampling time corresponds to that of Fig.~\ref{fig:trc_pdf}.
The grey dashed line corresponds to a Gaussian with a zero mean and unit variance.
}
\label{fig:single_struct_case}
\end{figure}

Fig.\ref{fig:pdf_fields_single} presents standard PDFs of streamwise and spanwise velocity fluctuations. In both cases, the PDFs are qualitatively similar, albeit slightly less so than for the early-time results shown in Fig.~\ref{fig:pd_pdfs_fields}.
A striking difference can be observed between the $u'$ PDFs in the space-filling and the single-structure state by comparison of the left panels of Figs.~\ref{fig:pd_pdfs_fields} and \ref{fig:pdf_fields_single}. While $u'$ is Gaussian distributed in the space-filling case, in the single-structure case the PDF is clearly bimodal and has non-zero skewness. In contrast, the spanwise velocity component PDF in the single-structure case shows similar slightly super-Gaussian tails as in the space-filling case, as can be seen by comparison of the right panels of Figs.~\ref{fig:pd_pdfs_fields} and \ref{fig:pdf_fields_single}. 
The physical origin of the deviation from Gaussian statistics for the streamwise velocity fluctuations in the single-structure state, as opposed to the space-filling state, will be investigated in future work.

Longitudinal and transverse velocity-gradient PDFs obtained from A1 and A6 for the single-structure state are presented in Fig.~\ref{fig:single_struct_case}. Apart from the PDFs for the A1-data having consistently noisier tails than the higher-resolved A6-data with positive definite polymer stresses, there is no significant difference between the two cases. Compared to the space-filling state shown in Fig.~\ref{fig:pd_pdfs_grad}, both longitudinal and transverse $w'-$ gradients fluctuate similarly in the single-structure state. This also applies to the longitudinal gradient of the streamwise velocity component. A different between the space-filling and the single-structure state can only be discerned for the transverse $u'-$gradient, whose PDF has an approximately Gaussian core in the space-filling state that is not present in the single-structure case.

\section{Conclusions}

The results presented above indicate that the behaviour of direct numerical simulations of elastic turbulence in channel flows is governed by two distinct spatial resolution thresholds. 
The first threshold corresponds to the minimum resolution required for stability thus yielding simulations that do not exhibit numerical, finite-time divergences. For the fluid parameters used in this study, this threshold is represented by the A1 run (see Table \ref{tab:datasets}). The second threshold corresponds to the minimum resolution necessary to ensure physically admissible values of the conformation tensor at all times throughout the domain. In our case, this condition is satisfied only by the highest-resolution simulation, A6.

Simulations performed at intermediate resolutions are numerically stable yet formally unphysical, as they fail to satisfy the condition $\mathrm{Tr}\, \c>3$ pointwise at all times. 
Remarkably, here we showed that such simulations, runs A1 through A5, still provide an accurate description of elastic turbulence. Key observables, including the statistics of velocity and velocity-gradient fluctuations, quantities previously used to characterise elastic turbulence in both experiments \citep{Groisman2000, Burghelea2007} and direct numerical simulations \citep{Dubief2013}, result in PDFs that are virtually indistinguishable from those obtained in the high-resolution A6 run, which strictly preserved the physical nature of the conformation tensor.

These findings have significant implications for direct numerical simulations of chaotic flows in polymeric liquids, particularly in wall-bounded shear flows. 
To date, there exists only a limited number of successful simulations of such flows (e.g. \cite{Zhang2013a,Dubief2013,Lellep2024,FoggiRota2024, Morozov2025preprint}) largely due to the extremely high spatial resolutions required to ensure the physical nature of the conformation tensor. 
For instance, the A6 run presented here took $1.6\cdot10^6$ core hours to complete, requiring access to a Tier 1 High-Performance Computing facility. 
In contrast, the A1 run completed using only $2\cdot10^5$ core hours - a computational cost attainable on most modern computing clusters.
Given that the A1 run yields flow structures and turbulent statistics that are virtually indistinguishable from the physically accurate A6 case, we propose that such low-resolution, formally unphysical simulations can serve as an efficient and practical tool for studying chaotic flows in polymeric liquids.

Alternatively, our findings support a hybrid-resolution strategy, motivated by the observation that the A1 and A6 runs sampled similar statistically-steady flow states. 
This approach would entail conducting a long, low-resolution run followed by multiple short high-resolution simulations initialised from snapshots along this trajectory. 
Our results suggest that only a few timesteps are needed in the high-resolution runs to eliminate unphysical regions in the conformation tensor, thereby producing a sequence of fully physical configurations suitable for further analysis.

The existence of a typical resolution required to preserve the physical nature of the polymeric stress implies the existence of an emergent microscopic lengthscale that needs to be  resolved for a direct numerical simulation to be admissible, similar to the Kolmogorov scale in Newtonian turbulence. 
While the origin of this lengthscale in elastic turbulence is currently unknown, here we speculate that it is related to the width of the stress filament that outlines the `body' and the `tusk' of the narwhal state, as discussed in \cite{Morozov2025preprint}. We defer the study of this lengthscale to our future work. 

In conclusion, we hope that our results will encourage a broader range of studies on elastic turbulence through direct numerical simulations, particularly those using (pseudo-)spectral methods. The computational strategies discussed above should help to lower the entry barrier to this field and pave the way for a wider exploration of elastic turbulence and its mechanism.

\section*{Acknowledgement.} 

The authors would like to thank EPSRC and UKRI for the computational time made available on the UK supercomputing facility ARCHER2 via the UK Turbulence Consortium (EP/X035484/1) and the `Access to High Performance Computing Facilities` call (APP95424). Support from the ARCHER2 team (https://www.archer2.ac.uk) are gratefully acknowledged. 

For the purpose of open access, the author has applied a Creative Commons Attribution (CC BY) licence to any Author Accepted Manuscript version arising from this submission.

\section*{Declaration of interests.} The authors report no conflicts of interest.

\bibliographystyle{jfm}
\bibliography{et}

\begin{thebibliography}{66}
\expandafter\ifx\csname natexlab\endcsname\relax\def\natexlab#1{#1}\fi
\def\au#1{#1} \def\ed#1{#1} \def\yr#1{#1}\def\at#1{#1}\def\jt#1{\textit{#1}}
  \def\bt#1{#1}\def\bvol#1{\textbf{#1}} \def\vol#1{#1} \def\pg#1{#1}
  \def\publ#1{#1}\def\arxiv#1{#1}\def\org#1{#1}\def\st#1{\textit{#1}}

\bibitem[Alves {\em et~al.\/}(2021)Alves, Oliveira \& Pinho]{Alves2021}
{\sc \au{Alves, MA}, \au{Oliveira, PJ} \& \au{Pinho, FT}} \yr{2021}
  \at{Numerical methods for viscoelastic fluid flows}.  \jt{Annu. Rev. Fluid
  Mech.}  \bvol{53},  \pg{509--541}.

\bibitem[Balci {\em et~al.\/}(2011)Balci, Thomases, Renardy \&
  Doering]{Balci2011}
{\sc \au{Balci, Nusret}, \au{Thomases, Becca}, \au{Renardy, Michael} \&
  \au{Doering, Charles~R.}} \yr{2011}  \at{Symmetric factorization of the
  conformation tensor in viscoelastic fluid models}.  \jt{J. Non-Newtonian
  Fluid Mech.}  \bvol{166},  \pg{546--553}.

\bibitem[Beneitez {\em et~al.\/}(2025)Beneitez, Mrini \&
  Kerswell]{Beneitez2025}
{\sc \au{Beneitez, Miguel}, \au{Mrini, Soufiane} \& \au{Kerswell, Rich~R.}}
  \yr{2025}  \at{Linear instability in planar viscoelastic {T}aylor–{C}ouette
  flow with and without explicit polymer diffusion}.  \jt{J. Non-Newtonian
  Fluid Mech.}  \bvol{345},  \pg{105459}.

\bibitem[Beneitez {\em et~al.\/}(2024{\natexlab{{\em a\/}}})Beneitez, Page,
  Dubief \& Kerswell]{Beneitez2024}
{\sc \au{Beneitez, Miguel}, \au{Page, Jacob}, \au{Dubief, Yves} \&
  \au{Kerswell, Rich~R.}} \yr{2024{\natexlab{{\em a\/}}}}  \at{Multistability
  of elasto-inertial two-dimensional channel flow}.  \jt{J. Fluid Mech.}
  \bvol{981},  \pg{A30}.

\bibitem[Beneitez {\em et~al.\/}(2024{\natexlab{{\em b\/}}})Beneitez, Page,
  Dubief \& Kerswell]{Beneitez2024a}
{\sc \au{Beneitez, M.}, \au{Page, J.}, \au{Dubief, Y.} \& \au{Kerswell, R.~R.}}
  \yr{2024{\natexlab{{\em b\/}}}}  \at{Transition route to elastic and
  elasto-inertial turbulence in polymer channel flows}.  \jt{Phys. Rev. Fluids}
   \bvol{9},  \pg{123302}.

\bibitem[Beneitez {\em et~al.\/}(2023)Beneitez, Page \& Kerswell]{Beneitez2023}
{\sc \au{Beneitez, Miguel}, \au{Page, Jacob} \& \au{Kerswell, Rich~R.}}
  \yr{2023}  \at{Polymer diffusive instability leading to elastic turbulence in
  plane {C}ouette flow}.  \jt{Phys. Rev. Fluids}  \bvol{8},  \pg{L101901}.

\bibitem[Berti {\em et~al.\/}(2008)Berti, Bistagnino, Boffetta, Celani \&
  Musacchio]{Berti2008}
{\sc \au{Berti, S.}, \au{Bistagnino, A.}, \au{Boffetta, G.}, \au{Celani, A.} \&
  \au{Musacchio, S.}} \yr{2008}  \at{Two-dimensional elastic turbulence}.
  \jt{Phys. Rev. E}  \bvol{77},  \pg{055306}.

\bibitem[Berti \& Boffetta(2010)]{Berti2010}
{\sc \au{Berti, S.} \& \au{Boffetta, G.}} \yr{2010}  \at{Elastic waves and
  transition to elastic turbulence in a two-dimensional viscoelastic
  {K}olmogorov flow}.  \jt{Phys. Rev. E}  \bvol{82},  \pg{036314}.

\bibitem[Bird {\em et~al.\/}(1987)Bird, Curtiss, Armstrong \&
  Hassager]{Bird1987_2}
{\sc \au{Bird, R.~B.}, \au{Curtiss, C.~F.}, \au{Armstrong, R.~C.} \&
  \au{Hassager, O.}} \yr{1987} {\em Dynamics of polymeric liquids\/}, 2nd edn.,
  ,  \vol{vol. 2. Kinetic theory}.  \publ{New York: Wiley}.

\bibitem[Burghelea {\em et~al.\/}(2007)Burghelea, Segre \&
  Steinberg]{Burghelea2007}
{\sc \au{Burghelea, Teodor}, \au{Segre, Enrico} \& \au{Steinberg, Victor}}
  \yr{2007}  \at{Elastic turbulence in von {K}arman swirling flow between two
  disks}.  \jt{Phys. Fluids}  \bvol{19},  \pg{053104}.

\bibitem[{Burns} {\em et~al.\/}(2020){Burns}, {Vasil}, {Oishi}, {Lecoanet} \&
  {Brown}]{Burns2020}
{\sc \au{{Burns}, Keaton~J.}, \au{{Vasil}, Geoffrey~M.}, \au{{Oishi},
  Jeffrey~S.}, \au{{Lecoanet}, Daniel} \& \au{{Brown}, Benjamin~P.}} \yr{2020}
  \at{{Dedalus: A flexible framework for numerical simulations with spectral
  methods}}.  \jt{Phys. Rev. Res.}  \bvol{2},  \pg{023068}.

\bibitem[Buza {\em et~al.\/}(2022)Buza, Beneitez, Page \& Kerswell]{Buza2022a}
{\sc \au{Buza, Gergely}, \au{Beneitez, Miguel}, \au{Page, Jacob} \&
  \au{Kerswell, Rich~R.}} \yr{2022}  \at{Finite-amplitude elastic waves in
  viscoelastic channel flow from large to zero {R}eynolds number}.  \jt{J.
  Fluid Mech.}  \bvol{951},  \pg{A3}.

\bibitem[{Castillo S\'{a}nchez} {\em et~al.\/}(2022){Castillo S\'{a}nchez},
  Jovanovi\'{c}, Kumar, Morozov, Shankar, Subramanian \& Wilson]{Sanchez2022}
{\sc \au{{Castillo S\'{a}nchez}, Hugo~A.}, \au{Jovanovi\'{c}, Mihailo~R.},
  \au{Kumar, Satish}, \au{Morozov, Alexander}, \au{Shankar, V.},
  \au{Subramanian, Ganesh} \& \au{Wilson, Helen~J.}} \yr{2022}
  \at{Understanding viscoelastic flow instabilities: Oldroyd-{B} and beyond}.
  \jt{J. Non-Newtonian Fluid Mech.}  \bvol{302},  \pg{104742}.

\bibitem[Chaudhary {\em et~al.\/}(2019)Chaudhary, Garg, Shankar \&
  Subramanian]{Chaudhary2019}
{\sc \au{Chaudhary, Indresh}, \au{Garg, Piyush}, \au{Shankar, V} \&
  \au{Subramanian, Ganesh}} \yr{2019}  \at{Elasto-inertial wall mode
  instabilities in viscoelastic plane {P}oiseuille flow}.  \jt{J. Fluid Mech.}
  \bvol{881},  \pg{119--163}.

\bibitem[Chaudhary {\em et~al.\/}(2021)Chaudhary, Garg, Subramanian \&
  Shankar]{Chaudhary2021}
{\sc \au{Chaudhary, Indresh}, \au{Garg, Piyush}, \au{Subramanian, Ganesh} \&
  \au{Shankar, V.}} \yr{2021}  \at{Linear instability of viscoelastic pipe
  flow}.  \jt{J. of Fluid Mech.}  \bvol{908},  \pg{A11}.

\bibitem[Choueiri {\em et~al.\/}(2018)Choueiri, Lopez \& Hof]{Choueiri2018}
{\sc \au{Choueiri, George~H}, \au{Lopez, Jose~M} \& \au{Hof, Bj{\"o}rn}}
  \yr{2018}  \at{Exceeding the asymptotic limit of polymer drag reduction}.
  \jt{Phys. Rev. Lett.}  \bvol{120},  \pg{124501}.

\bibitem[Choueiri {\em et~al.\/}(2021)Choueiri, Lopez, Varshney, Sankar \&
  Hof]{Choueiri2021}
{\sc \au{Choueiri, George~H.}, \au{Lopez, Jose~M.}, \au{Varshney, Atul},
  \au{Sankar, Sarath} \& \au{Hof, Bj{\"o}rn}} \yr{2021}  \at{Experimental
  observation of the origin and structure of elastoinertial turbulence}.
  \jt{Proc. Natl. Acad. Sci. U.S.A.}  \bvol{118}~(45).

\bibitem[Couchman {\em et~al.\/}(2024)Couchman, Beneitez, Page \&
  Kerswell]{Couchman2024}
{\sc \au{Couchman, Miles~M.P.}, \au{Beneitez, Miguel}, \au{Page, Jacob} \&
  \au{Kerswell, Rich~R.}} \yr{2024}  \at{Inertial enhancement of the polymer
  diffusive instability}.  \jt{J. Fluid Mech.}  \bvol{981},  \pg{A2}.

\bibitem[Datta {\em et~al.\/}(2022)Datta, Ardekani, Arratia, Beris,
  Bischofberger, McKinley, Eggers, L\'opez-Aguilar, Fielding, Frishman, Graham,
  Guasto, Haward, Shen, Hormozi, Morozov, Poole, Shankar, Shaqfeh, Stark,
  Steinberg, Subramanian \& Stone]{Datta2022}
{\sc \au{Datta, Sujit~S.}, \au{Ardekani, Arezoo~M.}, \au{Arratia, Paulo~E.},
  \au{Beris, Antony~N.}, \au{Bischofberger, Irmgard}, \au{McKinley, Gareth~H.},
  \au{Eggers, Jens~G.}, \au{L\'opez-Aguilar, J.~Esteban}, \au{Fielding,
  Suzanne~M.}, \au{Frishman, Anna}, \au{Graham, Michael~D.}, \au{Guasto,
  Jeffrey~S.}, \au{Haward, Simon~J.}, \au{Shen, Amy~Q.}, \au{Hormozi, Sarah},
  \au{Morozov, Alexander}, \au{Poole, Robert~J.}, \au{Shankar, V.},
  \au{Shaqfeh, Eric S.~G.}, \au{Stark, Holger}, \au{Steinberg, Victor},
  \au{Subramanian, Ganesh} \& \au{Stone, Howard~A.}} \yr{2022}
  \at{Perspectives on viscoelastic flow instabilities and elastic turbulence}.
  \jt{Phys. Rev. Fluids}  \bvol{7},  \pg{080701}.

\bibitem[Dubief {\em et~al.\/}(2023)Dubief, Terrapon \& Hof]{Dubief2023}
{\sc \au{Dubief, Yves}, \au{Terrapon, Vincent~E} \& \au{Hof, Bj{\"o}rn}}
  \yr{2023}  \at{Elasto-inertial turbulence}.  \jt{Annu. Rev. Fluid Mech.}
  \bvol{55},  \pg{675--705}.

\bibitem[Dubief {\em et~al.\/}(2013)Dubief, Terrapon \& Soria]{Dubief2013}
{\sc \au{Dubief, Yves}, \au{Terrapon, Vincent~E.} \& \au{Soria, Julio}}
  \yr{2013}  \at{On the mechanism of elasto-inertial turbulence}.  \jt{Phys.
  Fluids}  \bvol{25}~(11),  \pg{110817}.

\bibitem[Dubief {\em et~al.\/}(2005)Dubief, Terrapon, White, Shaqfeh, Moin \&
  Lele]{Dubief2005}
{\sc \au{Dubief, Yves}, \au{Terrapon, Vincent~E}, \au{White, Christopher~M},
  \au{Shaqfeh, Eric~SG}, \au{Moin, Parviz} \& \au{Lele, Sanjiva~K}} \yr{2005}
  \at{New answers on the interaction between polymers and vortices in turbulent
  flows}.  \jt{Flow Turbul. Combust.}  \bvol{74},  \pg{311--329}.

\bibitem[Dzanic {\em et~al.\/}(2022{\natexlab{{\em a\/}}})Dzanic, From \&
  Sauret]{Dzanic2022c}
{\sc \au{Dzanic, V.}, \au{From, C.S.} \& \au{Sauret, E.}}
  \yr{2022{\natexlab{{\em a\/}}}}  \at{The effect of periodicity in the elastic
  turbulence regime}.  \jt{J. of Fluid Mech.}  \bvol{937},  \pg{A31}.

\bibitem[Dzanic {\em et~al.\/}(2022{\natexlab{{\em b\/}}})Dzanic, From \&
  Sauret]{Dzanic2022}
{\sc \au{Dzanic, V.}, \au{From, C.S.} \& \au{Sauret, E.}}
  \yr{2022{\natexlab{{\em b\/}}}}  \at{A hybrid lattice {B}oltzmann model for
  simulating viscoelastic instabilities}.  \jt{Comput. Fluids}  \bvol{235},
  \pg{105280}.

\bibitem[Fattal \& Kupferman(2004)]{Fattal2004}
{\sc \au{Fattal, Raanan} \& \au{Kupferman, Raz}} \yr{2004}  \at{Constitutive
  laws for the matrix-logarithm of the conformation tensor}.  \jt{J.
  Non-Newtonian Fluid Mech.}  \bvol{123},  \pg{281--285}.

\bibitem[Foggi~Rota {\em et~al.\/}(2024)Foggi~Rota, Amor, Le~Clainche \&
  Rosti]{FoggiRota2024}
{\sc \au{Foggi~Rota, Giulio}, \au{Amor, Christian}, \au{Le~Clainche, Soledad}
  \& \au{Rosti, Marco~Edoardo}} \yr{2024}  \at{Unified view of elastic and
  elasto-inertial turbulence in channel flows at low and moderate {R}eynolds
  numbers}.  \jt{Phys. Rev. Fluids}  \bvol{9},  \pg{L122602}.

\bibitem[Garg {\em et~al.\/}(2018)Garg, Chaudhary, Khalid, Shankar \&
  Subramanian]{Garg2018}
{\sc \au{Garg, Piyush}, \au{Chaudhary, Indresh}, \au{Khalid, Mohammad},
  \au{Shankar, V.} \& \au{Subramanian, Ganesh}} \yr{2018}  \at{Viscoelastic
  pipe flow is linearly unstable}.  \jt{Phys. Rev. Lett.}  \bvol{121},
  \pg{024502}.

\bibitem[Graham \& Floryan(2021)]{Graham2021}
{\sc \au{Graham, Michael~D.} \& \au{Floryan, Daniel}} \yr{2021}  \at{Exact
  coherent states and the nonlinear dynamics of wall-bounded turbulent flows}.
  \jt{Annu. Rev. Fluid Mech.}  \bvol{53}~(1),  \pg{227--253}.

\bibitem[Groisman \& Steinberg(2000)]{Groisman2000}
{\sc \au{Groisman, Alexander} \& \au{Steinberg, Victor}} \yr{2000}  \at{Elastic
  turbulence in a polymer solution flow}.  \jt{Nature}  \bvol{405}~(6782),
  \pg{53--55}.

\bibitem[Hu \& Leli{\`e}vre(2007)]{Hu2007}
{\sc \au{Hu, Dan} \& \au{Leli{\`e}vre, Tony}} \yr{2007}  \at{New entropy
  estimates for {O}ldroyd-{B} and related models}.  \jt{Commun. Math. Sci.}
  \bvol{5},  \pg{909}.

\bibitem[Hulsen(1988)]{Hulsen88}
{\sc \au{Hulsen, Martien~A.}} \yr{1988}  \at{Some properties and analytical
  expressions for plane flow of {L}eonov and {G}iesekus models}.  \jt{J.
  Non-Newtonian Fluid Mech.}  \bvol{30},  \pg{85--92}.

\bibitem[Hulsen(1990)]{Hulsen90}
{\sc \au{Hulsen, Martien~A.}} \yr{1990}  \at{A sufficient condition for a
  positive definite configuration tensor in differential models}.  \jt{J.
  Non-Newtonian Fluid Mech.}  \bvol{38},  \pg{93--100}.

\bibitem[Khalid {\em et~al.\/}(2021{\natexlab{{\em a\/}}})Khalid, Chaudhary,
  Garg, Shankar \& Subramanian]{Khalid2021a}
{\sc \au{Khalid, Mohammad}, \au{Chaudhary, Indresh}, \au{Garg, Piyush},
  \au{Shankar, V.} \& \au{Subramanian, Ganesh}} \yr{2021{\natexlab{{\em a\/}}}}
   \at{The centre-mode instability of viscoelastic plane {P}oiseuille flow}.
  \jt{J. Fluid Mech.}  \bvol{915},  \pg{A43}.

\bibitem[Khalid {\em et~al.\/}(2021{\natexlab{{\em b\/}}})Khalid, Shankar \&
  Subramanian]{Khalid2021}
{\sc \au{Khalid, Mohammad}, \au{Shankar, V.} \& \au{Subramanian, Ganesh}}
  \yr{2021{\natexlab{{\em b\/}}}}  \at{Continuous pathway between the
  elasto-inertial and elastic turbulent states in viscoelastic channel flow}.
  \jt{Phys. Rev. Lett.}  \bvol{127},  \pg{134502}.

\bibitem[King {\em et~al.\/}(2026)King, Broadley \& Beneitez]{King2026preprint}
{\sc \au{King, Jack R.~C.}, \au{Broadley, Henry~M.} \& \au{Beneitez, Miguel}}
  \yr{2026} Elasto-inertial transitions in viscoelastic flows through cylinder
  arrays,  \arxiv{arXiv: 2604.05892}.

\bibitem[Kurganov \& Tadmor(2000)]{Kurganov2000}
{\sc \au{Kurganov, Alexander} \& \au{Tadmor, Eitan}} \yr{2000}  \at{New
  high-resolution central schemes for nonlinear conservation laws and
  convection–diffusion equations}.  \jt{J. Comput. Phys.}  \bvol{160},
  \pg{241--282}.

\bibitem[Larson(1999)]{Larson1999}
{\sc \au{Larson, R.~G.}} \yr{1999} {\em The structure and rheology of complex
  fluids\/}.  \publ{Oxford University Press}.

\bibitem[Lellep {\em et~al.\/}(2023)Lellep, Linkmann \& Morozov]{Lellep2023}
{\sc \au{Lellep, Martin}, \au{Linkmann, Moritz} \& \au{Morozov, Alexander}}
  \yr{2023}  \at{Linear stability analysis of purely elastic travelling-wave
  solutions in pressure-driven channel flows}.  \jt{J. Fluid Mech.}
  \bvol{959},  \pg{R1}.

\bibitem[Lellep {\em et~al.\/}(2024)Lellep, Linkmann \& Morozov]{Lellep2024}
{\sc \au{Lellep, Martin}, \au{Linkmann, Moritz} \& \au{Morozov, Alexander}}
  \yr{2024}  \at{Purely elastic turbulence in pressure-driven channel flows}.
  \jt{Proc. Natl. Acad. Sci. U.S.A.}  \bvol{121}~(9),  \pg{e2318851121}.

\bibitem[Lewy \& Kerswell(2024)]{Lewy2024}
{\sc \au{Lewy, Theo} \& \au{Kerswell, Rich}} \yr{2024}  \at{The polymer
  diffusive instability in highly concentrated polymeric fluids}.  \jt{J.
  Non-Newtonian Fluid Mech.}  \bvol{326},  \pg{105212}.

\bibitem[Lewy \& Kerswell(2025)]{Lewy2025}
{\sc \au{Lewy, Theo} \& \au{Kerswell, Rich~R.}} \yr{2025}  \at{Revisiting
  two-dimensional viscoelastic {K}olmogorov flow: a centre-mode-driven
  transition}.  \jt{J. Fluid Mech.}  \bvol{1007},  \pg{A55}.

\bibitem[Lin {\em et~al.\/}(2025)Lin, Liao, Zhao, Liu, Lu \& Khomami]{Lin2025}
{\sc \au{Lin, Fenghui}, \au{Liao, Zi-Mo}, \au{Zhao, Zhiye}, \au{Liu, Nansheng},
  \au{Lu, Xi-Yun} \& \au{Khomami, Bamin}} \yr{2025}  \at{{GPU} acceleration of
  a hi-fidelity algorithm for direct numerical simulation of
  polymer-induced/modified turbulence}.  \jt{J. Non-Newtonian Fluid Mech.}
  \bvol{342},  \pg{105437}.

\bibitem[Lopez {\em et~al.\/}(2019)Lopez, Choueiri \& Hof]{Lopez2019}
{\sc \au{Lopez, Jose~M.}, \au{Choueiri, George~H.} \& \au{Hof, Bj{\"o}rn}}
  \yr{2019}  \at{Dynamics of viscoelastic pipe flow at low {R}eynolds numbers
  in the maximum drag reduction limit}.  \jt{J. Fluid Mech.}  \bvol{874},
  \pg{699–719}.

\bibitem[Lu \& Hof(2026)]{Lu2026preprint}
{\sc \au{Lu, Ziyin} \& \au{Hof, Björn}} \yr{2026} Multiple states of
  turbulence at vanishing inertia,  \arxiv{arXiv: 2602.23498}.

\bibitem[Min {\em et~al.\/}(2001)Min, Yoo \& Choi]{Min2001}
{\sc \au{Min, Taegee}, \au{Yoo, Jung~Yul} \& \au{Choi, Haecheon}} \yr{2001}
  \at{Effect of spatial discretization schemes on numerical solutions of
  viscoelastic fluid flows}.  \jt{J. Non-Newtonian Fluid Mech.}  \bvol{100},
  \pg{27--47}.

\bibitem[Morozov(2022)]{Morozov2022}
{\sc \au{Morozov, Alexander}} \yr{2022}  \at{Coherent structures in plane
  channel flow of dilute polymer solutions with vanishing inertia}.  \jt{Phys.
  Rev. Lett.}  \bvol{129},  \pg{017801}.

\bibitem[Morozov {\em et~al.\/}(2025)Morozov, Lellep, Capocci \&
  Linkmann]{Morozov2025preprint}
{\sc \au{Morozov, Alexander}, \au{Lellep, Martin}, \au{Capocci, Damiano} \&
  \au{Linkmann, Moritz}} \yr{2025} Narwhals and their blessings: exact coherent
  structures of elastic turbulence in channel flows,  \arxiv{arXiv:
  2509.03175}.

\bibitem[Morozov \& Spagnolie(2015)]{Morozov2015}
{\sc \au{Morozov, Alexander} \& \au{Spagnolie, Saverio~E}} \yr{2015}
  \at{Introduction to complex fluids}.  \bt{In {\em Complex fluids in
  biological systems\/}},  \pg{pp. 3--52}.  \publ{Springer}.

\bibitem[Musacchio(2003)]{Musacchio2003}
{\sc \au{Musacchio, Stefano}} \yr{2003}  \at{Effects of friction and polymers
  on {2D} turbulence}. Phd thesis, Universit{\'a} degli Studi di Torino,
  Torino, Italy.

\bibitem[Nichols {\em et~al.\/}(2025)Nichols, Guy \& Thomases]{Nichols2025}
{\sc \au{Nichols, Jeffrey}, \au{Guy, Robert~D.} \& \au{Thomases, Becca}}
  \yr{2025}  \at{Period-doubling route to chaos in viscoelastic {K}olmogorov
  flow}.  \jt{Phys. Rev. Fluids}  \bvol{10},  \pg{L041301}.

\bibitem[Owens \& Phillips(2002)]{Owens2002}
{\sc \au{Owens, Robert~G} \& \au{Phillips, Timothy~N}} \yr{2002} {\em
  Computational rheology\/}.  \publ{World Scientific}.

\bibitem[Page {\em et~al.\/}(2020)Page, Dubief \& Kerswell]{Page2020}
{\sc \au{Page, Jacob}, \au{Dubief, Yves} \& \au{Kerswell, Rich~R.}} \yr{2020}
  \at{Exact traveling wave solutions in viscoelastic channel flow}.  \jt{Phys.
  Rev. Lett.}  \bvol{125},  \pg{154501}.

\bibitem[Phan-Thien \& Tanner(1977)]{PhanThien1977}
{\sc \au{Phan-Thien, Nhan} \& \au{Tanner, Roger~I.}} \yr{1977}  \at{A new
  constitutive equation derived from network theory}.  \jt{J. Non-Newtonian
  Fluid Mech.}  \bvol{2},  \pg{353--365}.

\bibitem[Samanta {\em et~al.\/}(2013)Samanta, Dubief, Holzner, Schäfer,
  Morozov, Wagner \& Hof]{Samanta2013}
{\sc \au{Samanta, Devranjan}, \au{Dubief, Yves}, \au{Holzner, Markus},
  \au{Schäfer, Christof}, \au{Morozov, Alexander~N.}, \au{Wagner, Christian}
  \& \au{Hof, Björn}} \yr{2013}  \at{Elasto-inertial turbulence}.  \jt{Proc.
  Natl. Acad. Sci. U.S.A.}  \bvol{110}~(26),  \pg{10557–10562}.

\bibitem[Shekar {\em et~al.\/}(2020)Shekar, McMullen, McKeon \&
  Graham]{Shekar2020}
{\sc \au{Shekar, Ashwin}, \au{McMullen, Ryan~M.}, \au{McKeon, Beverley~J.} \&
  \au{Graham, Michael~D.}} \yr{2020}  \at{Self-sustained elastoinertial
  {T}ollmien–{S}chlichting waves}.  \jt{J. Fluid Mech.}  \bvol{897},
  \pg{A3}.

\bibitem[Shekar {\em et~al.\/}(2021)Shekar, McMullen, McKeon \&
  Graham]{Shekar2021}
{\sc \au{Shekar, Ashwin}, \au{McMullen, Ryan~M.}, \au{McKeon, Beverley~J.} \&
  \au{Graham, Michael~D.}} \yr{2021}  \at{{T}ollmien-{S}chlichting route to
  elastoinertial turbulence in channel flow}.  \jt{Phys. Rev. Fluids}
  \bvol{6},  \pg{093301}.

\bibitem[Shekar {\em et~al.\/}(2019)Shekar, McMullen, Wang, McKeon \&
  Graham]{Shekar2019}
{\sc \au{Shekar, Ashwin}, \au{McMullen, Ryan~M.}, \au{Wang, Sung-Ning},
  \au{McKeon, Beverley~J.} \& \au{Graham, Michael~D.}} \yr{2019}
  \at{Critical-layer structures and mechanisms in elastoinertial turbulence}.
  \jt{Phys. Rev. Lett.}  \bvol{122},  \pg{124503}.

\bibitem[Sid {\em et~al.\/}(2018)Sid, Terrapon \& Dubief]{Sid2018}
{\sc \au{Sid, S.}, \au{Terrapon, V.~E.} \& \au{Dubief, Y.}} \yr{2018}
  \at{Two-dimensional dynamics of elasto-inertial turbulence and its role in
  polymer drag reduction}.  \jt{Phys. Rev. Fluids}  \bvol{3},  \pg{011301}.

\bibitem[Steinberg(2021)]{Steinberg2021}
{\sc \au{Steinberg, Victor}} \yr{2021}  \at{Elastic turbulence: an experimental
  view on inertialess random flow}.  \jt{Annu. Rev. Fluid Mech.}  \bvol{53},
  \pg{27--58}.

\bibitem[Steinberg(2022)]{Steinberg2022}
{\sc \au{Steinberg, Victor}} \yr{2022}  \at{New direction and perspectives in
  elastic instability and turbulence in various viscoelastic flow geometries
  without inertia}.  \jt{Low Temp. Phys.}  \bvol{48},  \pg{492--507}.

\bibitem[Thomas {\em et~al.\/}(2006)Thomas, Al-Mubaiyedh, Sureshkumar \&
  Khomami]{Thomas2006}
{\sc \au{Thomas, D.G.}, \au{Al-Mubaiyedh, U.A.}, \au{Sureshkumar, R.} \&
  \au{Khomami, B.}} \yr{2006}  \at{Time-dependent simulations of
  non-axisymmetric patterns in {T}aylor–{C}ouette flow of dilute polymer
  solutions}.  \jt{J. Non-Newton. Fluid Mech.}  \bvol{138},  \pg{111--133}.

\bibitem[Vaithianathan \& Collins(2003)]{Vaithianathan2003}
{\sc \au{Vaithianathan, T.} \& \au{Collins, Lance~R.}} \yr{2003}  \at{Numerical
  approach to simulating turbulent flow of a viscoelastic polymer solution}.
  \jt{JJ. Comput. Phys.}  \bvol{187},  \pg{1--21}.

\bibitem[Wang \& Ruuth(2008)]{Wang2008}
{\sc \au{Wang, Dong} \& \au{Ruuth, Steven~J}} \yr{2008}  \at{Variable step-size
  implicit-explicit linear multistep methods for time-dependent partial
  differential equations}.  \jt{J. Comput. Math.}  \bvol{26},  \pg{838--855}.

\bibitem[Yerasi {\em et~al.\/}(2024)Yerasi, Picardo, Gupta \&
  Vincenzi]{Yerasi2024}
{\sc \au{Yerasi, Sumithra~R}, \au{Picardo, Jason~R}, \au{Gupta, Anupam} \&
  \au{Vincenzi, Dario}} \yr{2024}  \at{Preserving large-scale features in
  simulations of elastic turbulence}.  \jt{J. Fluid Mech.}  \bvol{1000},
  \pg{A37}.

\bibitem[Zhang {\em et~al.\/}(2013)Zhang, Li, Cao, Tomoaki \& Yu]{Zhang2013a}
{\sc \au{Zhang, Hong-Na}, \au{Li, Feng-Chen}, \au{Cao, Yang}, \au{Tomoaki,
  Kunugi} \& \au{Yu, Bo}} \yr{2013}  \at{Direct numerical simulation of elastic
  turbulence and its mixing-enhancement effect in a straight channel flow}.
  \jt{Chin. Phys. B}  \bvol{22},  \pg{024703}.

\bibitem[Zhu \& Kerswell(2024)]{Zhu2024preprint}
{\sc \au{Zhu, Lu} \& \au{Kerswell, Rich~R.}} \yr{2024} Early turbulence in
  viscoelastic flow past a periodic cylinder array,  \arxiv{arXiv: 2410.12033}.

\end{thebibliography}

\end{document}